\documentclass[toc]{PoS}

\usepackage{pdflscape}
\usepackage{longtable}
\usepackage{amssymb}
\usepackage{amsmath}
\usepackage{adjustbox}
\usepackage{threeparttable}
\usepackage{tabto}
\usepackage{float}
\usepackage{color}
\usepackage{graphics}
\usepackage{multirow}
\usepackage{multicol}

\newcommand{\arcmin}{\hbox{$^\prime$}}
\newcommand{\arcsec}{\hbox{$^{\prime\prime}$}}
\newcommand{\Msun}{M$_{\odot}$}

\definecolor{smalt(darkpowderblue)}{rgb}{0.0, 0.2, 0.6}
\definecolor{forestgreen(traditional)}{rgb}{0.0, 0.5, 0.0}

\newcommand{\caml}{consequential angular momentum loss}

\newcommand{\ibp}{initial binary population}

\newcommand{\msto}{MS stars close to the turn-off point}
\newcommand{\bse}{{\sc bse}}
\newcommand{\mocca}{{\sc mocca}}

\newcommand{\hst}{\emph{HST}}
\newcommand{\chandra}{\emph{Chandra}}

\newcommand{\muse}{MUSE}
\newcommand{\gaia}{\emph{Gaia}}
\newcommand{\sdss}{\emph{Sloan Digital Sky Survey}}

\title{Properties of 
Cataclysmic Variables 
in Globular Clusters}

\ShortTitle{Properties of 
CVs in GCs}

\author{\speaker{Diogo Belloni}\\
       National Institute for Space Research, 
       Sao Jose dos Campos, Brazil\\
       E-mail: \email{diogo.belloni@inpe.br}}

\author{Liliana E. Rivera Sandoval\\
       Texas Tech University, Lubbock, USA\\
       E-mail: \email{Liliana.Rivera@ttu.edu}}

\abstract{The study of star clusters plays an important role in our understanding of the Universe since these systems are natural laboratories for testing theories of stellar dynamics and evolution. Particularly, globular clusters (GCs) are one of the most important objects for studying the formation and the physical nature of exotic systems
which in turn provide basic information and tools that can help us to understand the formation and evolution processes of star clusters themselves, galaxies and, in general, the young Universe. Among the most interesting objects in GCs are the cataclysmic variables (CVs), which are interacting binaries harboring a white dwarf accreting from a low-mass companion. Since GC densities are sufficiently high that dynamical encounters involving binaries should be common, CV progenitors are expected to be affected by dynamics in some way 
in the early stages. In this article we review the formation channels and the influence of dynamics on the CV population in GCs. In particular, we review recent progress in numerical simulations. Furthermore, we discuss observational properties of CVs in GCs and the techniques used to identify and study them. We focus the discussion on the multi-wavelength observations carried out with \hst~ and \chandra~ on the best-studied GCs NGC~6397, NGC~6752, 47~Tucanae and $\omega$~Centauri; on the recent spectroscopic findings with \muse, and on updates regarding the correlation between the number of faint X-ray sources and the cluster stellar encounter rate. Finally, we discuss some observational prospects that might potentially help future investigations.}

\FullConference{The Golden Age of 
        Cataclysmic Variables 
        and Related Objects V\\
        2--7 September, 2019\\
        Palermo, Italy}

\begin{document}

\makeatletter
\setbox\@firstaubox\hbox{\small Diogo Belloni \& Liliana E. Rivera Sandoval}
\makeatother


\section{Why should one investigate cataclysmic variables in globular clusters?}
\label{SecINTRO}

The study of star clusters plays an important role in our understanding of the Universe since these systems are natural laboratories for testing theories of stellar dynamics and evolution.
Particularly, globular clusters (GCs)
are one of the most important objects for studying the formation and the physical nature of exotic systems given that they are nearly as old as the Universe itself and they can reach very high stellar densities (up to $\sim10^6$ stars per pc$^3$ in their cores, Harris 1996~\cite{Harris_1996}).
Because stellar dynamics plays an important role in these environments, GCs are factories of exotic systems such as low mass X-ray binaries, cataclysmic variables, active binaries, red and blue straggler stars, millisecond pulsars, black hole binaries, among others.
Thus, the study of these binaries provides key information and tools that can help us to understand the formation and evolution processes of star clusters themselves, which, in turn, help us to understand galaxies and, in general, the young Universe (e.g. Spitzer 1987~\cite{Spitzer_BOOK}; Hut et al. 1992~\cite{Hut_1992}; Meylan \& Heggie 1997~\cite{Meylan_1997}; Ashman \& Zepf 1998~\cite{Ashman_1998}; Heggie \& Hut 2003~\cite{HeggieBOOK}; Brodie \& Strader 2006~\cite{Brodie_2006}; Aarseth, Tout \& Mardling 2008~\cite{Aarseth_2008}; Binney \& Tremaine 2008~\cite{Binney_2008}; Portegies Zwart, McMillan \& Gieles 2010~\cite{Portegies_2010}; Davies 2013~\cite{Davies_2013}; Bastian \& Lardo 2018~\cite{Bastian_2018}; Varri et al. 2018~\cite{Varri_2018}; Baumgardt et al. 2019~\cite{Baumgardt_2019}; Gratton et al. 2019~\cite{Gratton_2019}).

Among the most interesting objects in GCs are the cataclysmic variables (CVs),
which are interacting binaries composed of a white dwarf (WD) that stably accretes matter from a low-mass main-sequence (MS) star (e.g. Smak 1984~\cite{Smak_1984}; Cropper 1990~\cite{Cropper_1990}; Warner 1995~\cite{Warner_1995_OK}; Smak 2000~\cite{Smak_2000}; Hellier 2001~\cite{Hellier_2001}; Lasota 2001~\cite{Lasota_2001}; Knigge, Baraffe \& Patterson 2011~\cite{Knigge_2011_OK}; Ferrario, de Martino \& G\"ansicke 2015~\cite{Ferrario_2015a}; Zorotovic \& Schreiber 2019~\cite{Zorotovic_2019}; Ferrario, Wickramasinghe \& Kawka 2020~\cite{Ferrario_2020}).
CVs are important astrophysical systems, since they are the most abundant class of interacting binaries harboring a compact object,
not only in the Milky Way field (Breedt [private communication]), but also in GCs (e.g. there are from 100 to $1000$~times more WDs than neutron stars in a GC, Maccarone \& Knigge 2007~\cite{Maccarone_2007}). 
Furthermore, due to their different types of variability (which basically occurs over all wavelengths and on a wide range of time-scales) caused by different physical processes, CVs become relevant in several fields of research.
In fact, besides allowing us to investigate close binary formation and evolution, CVs also allow us to get insights about the physics of accretion as well as the interactions between high-density plasma and magnetic fields. 

GCs are thought to play a crucial role in CV formation, since their densities are sufficiently high that dynamical encounters involving binaries should be common. 
Thus, in dense GCs, it is natural to expect that CVs exist in non-negligible numbers, and that many CV progenitors will have been affected by dynamics in some way prior to the CV formation (e.g. Bailyn, Grindlay \& Garcia 1990~\cite{Bailyn_1990}; Stefano \&  Rappaport 1994~\cite{Stefano_1994}; Davies 1995~\cite{Davies_1995}; Davies \& Benz 1995~\cite{Davies_1995b}; Davies 1997~\cite{Davies_1997}; Shara \& Hurley 2006~\cite{Shara_2006}; Ivanova et al. 2006~\cite{Ivanova_2006}; Hong et al. 2017~\cite{Hong_2017}; Belloni et al. 2016~\cite{Belloni_2016a}; 2017a~\cite{Belloni_2017a}; 2017b~\cite{Belloni_2017b}; 2019~\cite{Belloni_2019}).
Furthermore, there is a huge degeneracy regarding GC modelling, in the sense that, in a lot of cases, different evolutionary pathways associated with different initial models lead to comparable present-day global properties.
Such a degeneracy could in principle be broken by investigating particular types of objects, whose properties strongly depend on the environment to which they belong.
Therefore, some particular objects might potentially provide useful constrains on initial GC conditions and on the \ibp.
For example, while comparing predicted and observed binary fractions in GCs, Leigh et al. (2015~\cite{Leigh_2015}) were able to conclude that only high initial binary fractions (with a significant fraction of very wide binaries) combined with high initial densities can reproduce the observed anti-correlation between the binary fraction (both inside and outside the half-mass radius) and the total cluster mass.
Another example is the work by Belloni et al. (2017c~\cite{Belloni_2017c}), in which these authors compared predicted and observed properties of GCs based on the colour-magnitude diagrams (CMDs) and were able to further improve properties of initial binaries in order to incorporate the observations into the numerical simulation investigations.
In a similar fashion, comparing predicted and observed CV properties might lead to relevant constraints for the initial GC conditions and their initial binary characteristics.
One of the main problems (if not the major one) in GC modelling is how to deal with the unknown initial properties, and providing useful constraints can potentially help to reduce the degrees of freedom while modelling these systems.

In this document, we give an updated view of the recent progress in both, the theoretical and observational studies of GC CVs. We also discuss directions for future studies in order to increase our knowledge of the topic.
For a previous review on the main properties of GC CVs, see Knigge (2012~\cite{Knigge_2012MMSAI}).

\section{What do we know from observations?}
\label{SecOBS}

The very first CV observed in a GC was T Scorpii (in the core of M80), which is a classical nova
and was detected independently by the astronomers Arthur Auwers (Luther 1860~\cite{Luther_1860}) and Norman Pogson (Pogson 1860~\cite{Pogson_1860}), in the 19th century, and discussed in detail by Shara \& Drissen (1995~\cite{Shara_1995b}).
A more common type of CVs are the dwarf novae (DNe)
which constitute the majority of CVs in the Milky Way field (e.g. Pala et al. 2020~\cite{Pala_2020}). The very first DN identified in a GC was V101 in the outskirts of M5.
V101 was classified as an SS Cygni-like CV by Oosterhoff (1941~\cite{Oosterhoff_1941}), and its nature spectroscopically confirmed with ground-based observations by Margon, Downes \& Gunn (1981~\cite{Margon_1981}), who obtained a low resolution spectrum showing strong, broad Balmer and He I emission lines.
Margon \& Downes (1983~\cite{Margon_1983}) also claimed to had found the second DN in a GC (V4 in M30), but this DN is now thought to be a foreground object in the field of M30 (Machin et al. 1991~\cite{Machin_1991}; Kains et al. 2013~\cite{Kains_2013}), which is an example of reclassification of a presumable CV in a GC when obtaining more information. 
Except for V101, the first searches for DNe performed with ground-based telescopes proved to be very unsuccessful (e.g. Webbink 1980~\cite{Webbink_1980}). 
This trend has remained even with more sophisticated ground-based instruments and surveys (e.g. Kaluzny et al. 2005~\cite{Kaluzny_2005}; Pietrukowicz et al. 2008~\cite{Pietrukowicz_2008}). 
In fact, because of the low WD intrinsic luminosities and the high stellar crowding in GCs, it has been difficult for ground-based telescopes to reach the vast WD population in any GC. It was until the launch of the \emph{Hubble Space Telescope} (\hst), that with its extremely good spatial resolution and sensitivity, the very first detection of a WD cooling sequence in a GC was done (Richer et al. 1995~\cite{Richer_1995}).
However, the detection of DNe in GCs has been difficult even with space telescopes such as the \hst. 
As we will discuss later, the most effective way to look for CVs in GCs seems to be a combination of several types of data.

\subsection{Searches for cataclysmic variables in globular clusters}
\label{Sec2.1}

Since most CVs in the Milky Way are expected to be DNe (e.g. Pala et al. 2020~\cite{Pala_2020}), it is natural to think of searching for them based on the variability associated with DN outbursts.
The first attempts to detect DN outbursts in GCs were carried out by Shara et al. in a few GCs.
For example, Shara et al. (1988~\cite{Shara_1988}) searched for variability
in $\omega$~Cen and 47~Tuc, and found no DNe eruptions. 
Shara, Bergeron \& Moffat (1994~\cite{Shara_1994}) looked for erupting DNe outside the core of M92 and also found none.
Shara et al. (1995~\cite{Shara_1995a}) tried to identify variable sources in part of the core of NGC~6752 and also found none.
Shara \& Drissen (1995~\cite{Shara_1995b}) investigated the central regions of M80 and were able to only recover old novae, and in turn found no DNe.
However, Paresce et al. (1994~\cite{Paresce_1994}) identified the second DN in a GC (in 47~Tuc), which later was again identified in outburst by Shara et al. (1996~\cite{Shara_1996}) using 12 separate epochs of \hst \ images in visual through near-ultraviolet (NUV) bands.

The most recent and complete survey dedicated to search for DNe in GCs 
was performed by Pietrukowicz et al. (2008~\cite{Pietrukowicz_2008}).
It is based on the largest available homogeneous sample of observations, in terms of the time-span, number of observations and number of GCs (a total of 16 Galactic GCs).
These authors found no more than 12 DNe, distributed among seven clusters.
More recently, Modiano, Parikh \& Wijnands (2020~\cite{Modiano_2020}) confirmed 3 known CVs as DNe in the GC 47 Tuc, bringing the total number of confirmed DNe in all GCs to 17 (there are indications of a few more CVs to be DNe but these have not been firmly confirmed yet, see e.g. Rivera Sandoval et al. 2018~\cite{Rivera_2018}).

At first glance these works point towards the idea that there is a genuine deficiency of DNe in GCs and thus clearly conflicting with theoretical works, which predicts that most CVs should be DNe and that hundreds (and even thousands) of CVs exist in each GC.
However, one way of interpreting such results is translating it as a deficit of DN outbursts, not DNe necessarily.
This would suggest that, if the majority of GC CVs are indeed DNe, then their duty cycles should be extremely small.
In this case, alternative methods than variability through DN outbursts could help to identify GC CV candidates.
Another way of searching for CVs (not only DNe) in GCs is by taking into account that CVs are composed of accreting WDs, which means that looking for signatures of the accretion process could be an efficient method to detect them.
This is, in fact, the approach that has been followed by many authors since the \emph{Chandra X-ray Observatory} was launched, and which has allowed the discovery of many CVs thanks to its exquisite sensitivity and angular resolution.
Under such an approach, CV candidates are identified based on the detection of their X-ray emission (see e.g. Heinke 2010~\cite{Heinke_2010}, and references therein).

Accreting WDs show different characteristics from non-accreting WDs in their X-ray, UV, optical and H$\alpha$ emissions, given that the emission in most accreting systems is powered by the accretion process.
However, detecting signatures of accretion in period bouncers
seems to be extremely difficult (e.g. Hern\'andez Santisteban et al. 2018~\cite{Santisteban_2018}) because these systems are observed basically as single WDs.
Given that CVs are relatively hot systems, they are expected to be bluer than normal MS stars (or outliers) in GC CMDs (e.g. Campos et al. 2018~\cite{Campos_2018}).
Additionally, they should also reveal excess in UV and H$\alpha$ emissions (e.g. Cool et al., 1995 \cite{1995cool6397}; Knigge et al., 2008 \cite{Knigge_2008}; Beccari et al., 2014 \cite{2014beccari}).
Finally, they could also be detected as X-ray sources, which is generated in the boundary layer (in case of DNe) or in the post-shock region (in case of magnetic CVs) (see e.g. van den Berg 2020 \cite{vandenBerg_2020} and references therein).

The best approach then seems to be a technique that simultaneously combines data taken in different bands (X-ray, UV, optical, IR) in which information in narrow filters such as H$\alpha$ is included, and where variability is considered. That technique (as a whole) was firstly applied by Rivera Sandoval et al. (2018~\cite{Rivera_2018}) in 47 Tuc, which led to the discovery of 22 new CV systems. 

There are also many other GCs in which the \emph{multi-wavelength approach} (using data in at least 2 different bands) was adopted (since the 1980's). 
In most of these investigations, when X-ray data has been available, the authors have proceeded (summarily) as follows: they first identified X-rays sources in a GC (mainly with \chandra), and then searched for photometric counterparts of such sources (primarily with \hst, the only telescope capable to resolve the cores of GCs), mainly in UV, optical, NIR, and/or H$\alpha$ bands.  
In this way, in all these investigations,
the authors managed to identify CV candidates effectively through their position in the CMDs, the color-color diagrams and (in some cases) by identifying short-term variability. However, such method is of course X-ray biased.
We emphasise that UV data has tremendously helped to the identification of CVs in GCs, since such data cannot be obtained from the ground and it is where CVs mostly emit. 
Furthermore, since the stars in GCs are old and red, UV observations help to overcome the stellar crowding, thus revealing the faint emission of CVs (compared to other stars in the cluster).  
All these investigations confirm the picture that the multi-wavelength approach is more effective than searching for DN outbursts, provided the lack of data, in number, in filters, and with the appropriate cadences to do that.

The reader should have noticed by now the emphasis here employed to the term `\textit{CV candidate}' instead of simply `\textit{CV}'.
This is because, even though one considers all evidences at hand to support the idea that a source is a CV, as in the Milky Way field, spectroscopic confirmation is the only way to know for sure that a particular source is a CV (e.g. Knigge 2012~\cite{Knigge_2012MMSAI}).
As one would expect, this is a very difficult task given the faintness of CVs in GCs, coupled with the crowded nature of the GC environment.
For simplicity, hereafter we will not distinguish any more CV candidates from confirmed CVs, and refer to both groups simply as CVs. The reader should keep in mind, though, that the majority of the CVs in GCs are only candidates, lacking firm spectroscopic confirmation.

\subsection{Spectroscopic observations of cataclysmic variables in globular clusters}
\label{Sec4.2}

As in the Galactic field, the best way to identify the nature of a system is by obtaining spectroscopic observations. 
Following Knigge (2012~\cite{Knigge_2012MMSAI}), this method is the {\it gold standard} to confirm CVs, while the photometric methods are the {\it silver standard}. 
In the field, several surveys have led to the successful spectroscopic identification of CVs. For instance, the \sdss~provides both photometric and spectroscopic data, and has allowed the unambiguous identification of CVs from hydrogen and helium emission lines (see paper series by Szkody et al. 2002-2011~\cite{Szkody_2002,Szkody_2011} and references therein).

However, in GCs there are two major complications to identify CVs through spectroscopic observations, these are the stellar crowding and the intrinsic faintness of the systems.
This means that the same instruments that have been successful in identifying CVs from the ground are not applicable in GCs given the spatial resolution necessary to resolve the cores (where a large populations of CVs is expected to reside) and to avoid contamination from nearby stars.
These complications have led to a very limited number of CVs spectroscopically confirmed (17 CVs so far; G\"ottgens et al., 2019b \cite{Gottgens_2019b}).
One of the instruments that has led to half of these identifications is \hst, mainly using the long-slit technique. However, Knigge et al. (2008,~\cite{Knigge_2008}), also exploited the power of slitless spectroscopy in the FUV, where only the hottest objects emit. That approach led to the simultaneous confirmation of three systems in 47 Tuc.

Unfortunately, \hst \ spectroscopy is difficult to perform given the large number of orbits necessary to obtain spectra of faint CVs using long-slits.
Moreover, the field of view of the FUV detectors (to perform slitless spectroscopy) is small with respect to the cores of several globular clusters and thus, several pointings are required.
Additionally, several GCs have large extinction in that band which also makes difficult to perform FUV slitless spectroscopy, especially for faint systems.
However, in the last few years the development of more sophisticated ground-based spectrographs, which involve adaptive optics and larger fields of view (than FUV \hst \ instruments) have allowed the study of short period binaries in crowded environments such as GCs.
One of these instruments is the \emph{Multi Unit Spectroscopic Explorer} (\muse), which is a panoramic integral-field spectrograph working at optical wavelengths. It is located on the \emph{Very Large Telescope} in Chile (Bacon et al. 2010~\cite{Bacon_2010}).
With a field of view as large as $1\arcmin\times 1\arcmin$ (in the wide field mode) and a spatial resolution as small as $0.03\arcsec$ (in the narrow field mode), MUSE is able to cover the cores of GCs and study a large variety of systems in such environments.

Furthermore, the combination of catalogues obtained with \hst, together with a point spread function fitting technique, has allowed to deblend objects in crowded fields, thus increasing the power of MUSE in GCs (e.g. Kamann et al. 2013~\cite{Kamann_2013}).
This technique has been applied to several GCs and has led to the study of thousands of stars in such environments, e.g. NGC~6397 (Husser et al. 2016~\cite{Husser_2016}; Kamann et al. 2016~\cite{Kamann_2016}), NGC~2808 (Latour et al. 2019~\cite{Latour_2019}), NGC~3201 (Giesers et al. 2018~\cite{Giesers_2018}; 2019~\cite{Giesers_2019}), Terzan~9 (Ernandes et al. 2019~\cite{Ernandes_2019}), 47~Tuc (Kamann et al. 2018~\cite{Kamann_2018}).
More recently, MUSE has been used to observe 26 GCs to identify emission-line sources, which has led to the confirmation of five CVs and the discovery of two more (G\"ottgens et al. 2019b~\cite{Gottgens_2019b}).
This is important as previous to the MUSE era, only ten CVs had been spectroscopically confirmed in the whole Galactic GC population (Margon, Downes \& Gunn 1981~\cite{Margon_1981}, Grindlay et al. 1995~\cite{Grindlay_1995}, Edmonds et al. 1999~\cite{Edmonds_1999}, Deutsch et al. 1999~\cite{Deutsch_1999}, Knigge et al. 2003~\cite{Knigge_2003}, 2008~\cite{Knigge_2008}, Webb \& Servillat 2013~\cite{Webb_2013}).
However, the faint luminosities of the several photometrically identified CVs make such systems very difficult to detect by MUSE. 
For example, in cases like 47~Tuc, where most of the CVs have been identified only in the UV or at optical luminosities larger than 22~mag (Rivera Sandoval et al. 2018~\cite{Rivera_2018}), MUSE is unable to detect them (those around 26~mag) because they are below its detection limit.
Though, the detection of some of the faint systems can be performed with long exposure times, and by using the narrow field mode, which has a better spatial resolution.

The discussion so far leads us to the conclusion that {\it future observations which objective is to find CVs should proceed by means of the multi-wavelength approach} provided that it is easier to identify faint systems with it, though an X-ray unbiased search would be essential in order to make direct comparisons with CVs in the Milky Way field (Rivera Sandoval et al. [in preparation]). 
We already discussed that observational surveys trying to identify DNe via their variability through outbursts have been mostly unsuccessful given the filters and cadences used. However, that technique is very promising if carried out in UV and with a relatively short cadence (a few days or perhaps even a few weeks between observations) and with a relatively long monitoring program (longer than a few days or a few weeks). A good spatial resolution is then critical for such searches in order to resolve the cores of GCs, and this can only be achieved so far with \hst \ in UV, though blue filters using adaptive optics can also be helpful.
We stress that searches using only one band could lead to unreliable or to an incomplete identification of the CV population (e.g. some X-ray sources that were classified initially as CVs have been later confirmed as other type of exotic binaries when adding information in other bands).
On the other hand, the multi-wavelength approach has already been proved a successful method, as discussed previously. It should then keep being adopted in future observations, as it seems to be the best alternative our current instrumental resources provide us.
However, in order to accurately understand the GC CV population, spectroscopic confirmation and characterisation, such as determining the masses of the components, the mass transfer rate as well as the orbital periods are very important.

\subsection{Particular cases: NGC~6397, NGC~6752, $\omega$~Cen and 47~Tuc}
\label{Sec2.2}

We now focus on the more extensively studied (with a multi-wavelength approach) GCs with respect to CV populations, namely NGC~6397, NGC~6752, $\omega$~Cen, and 47~Tuc because they harbor the large majority of the CVs we know in the Galactic GCs (there are a bit more than 100 CVs in these four clusters).

\subsubsection{Optical and X-ray properties}
\label{Sec2.2.1}

\begin{figure}[t]
  \begin{center}

    \includegraphics[width=0.48\linewidth]{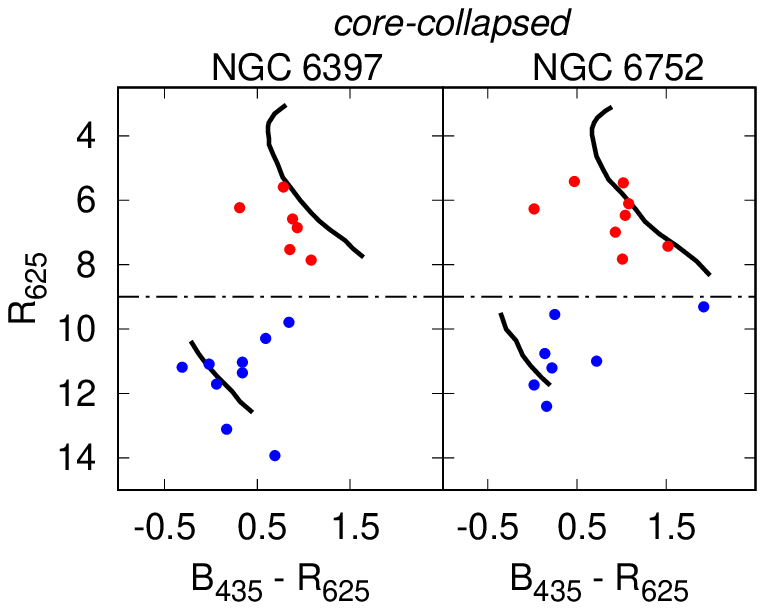}
    \hspace{0.2cm}
    \includegraphics[width=0.48\linewidth]{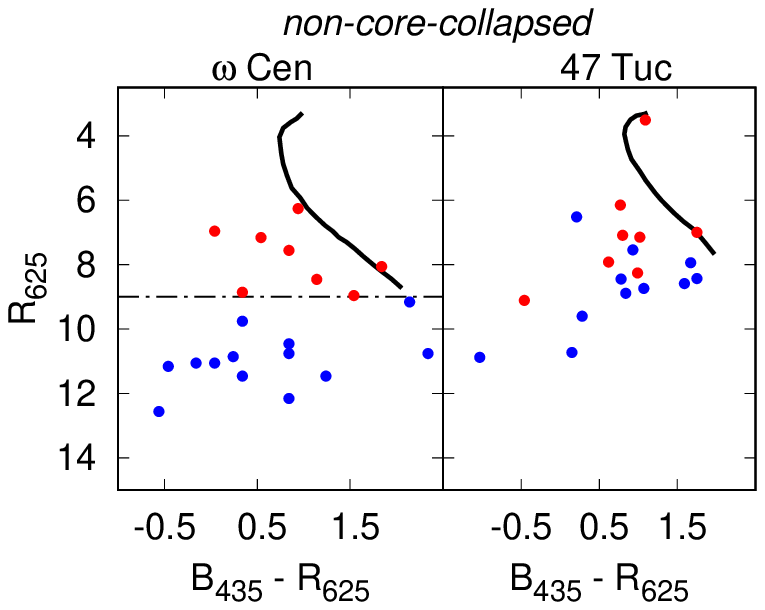}
  \end{center}
  \caption{CMDs for MS stars and WD cooling sequence (black lines) within the half-light radius of NGC~6397, NGC~6752, $\omega$~Cen, and 47~Tuc, along with CV identifications (red and blue points are bright and faint systems, respectively). It is clear that core-collapsed GCs have two populations, one being brighter than the other, while the same is not easily observed for non-core-collapsed GCs. To distinguish both populations in each cluster, we used a cut-off of ${R_{\rm 625}\simeq9}$, for core-collapsed clusters and $\omega$~Cen (horizontal dashed lines), and ${B_{\rm 390}\simeq9}$, for 47~Tuc given that most systems there have been identified in bluer bands. Panels produced with data from Cohn et al. (2010~\cite{Cohn_2010}; NGC~6397), Lugger et al. (2017~\cite{Lugger_2017}; NGC~6752), Cool et al. (2013~\cite{Cool_2013}; $\omega$~Cen), and Rivera Sandoval et al. (2018~\cite{Rivera_2018}; 47~Tuc). Absolute magnitudes were computed 
  assuming the following distances (in kpc): 2.3 (for NGC~6397); 4.0 (for NGC~6752); 5.2 (for $\omega$~Cen); 4.69 kpc (for 47~Tuc). CMDs are corrected by extinction and X-ray sources with ambiguous counterparts were not included.}
  \label{FigCMDobs}
\end{figure}

Let us start then our discussion by presenting results based on one of the most important tools in astrophysical research, i.e. CMDs.
By inspecting the CMDs in NGC~6397 (Cohn et al. 2010~\cite{Cohn_2010}), NGC~6752 (lugger et al. 2017~\cite{Lugger_2017}), $\omega$~Cen (Cool et al. 2013~\cite{Cool_2013}) and 47~Tuc (Rivera Sandoval et al. 2018~\cite{Rivera_2018}) showed in Fig. \ref{FigCMDobs}, we can see a very interesting difference between core-collapsed and non-core-collapsed clusters.
In core-collapsed clusters (NGC~6397 and NGC~6752), we can clearly identify two populations of CVs, one brighter than the other, the cut-off being at ${R_{625}\simeq9}$~mag.
The {\it bright} CVs in both clusters lie close to the MS stars and the {\it faint} CVs lie close to the WD cooling sequence.
All these studies included the same \hst \ filters, here named B$_{435}$ and R$_{625}$. 
Since CVs above the orbital period gap have more massive donors (Knigge et al. 2006 \cite{Knigge_2006}), they mostly emit in the R$_{625}$ band, which suggests that bright CVs are likely above the orbital period gap and harbor more massive donors ($\gtrsim0.5$~\Msun). 
Regarding faint CVs, their fluxes are usually dominated by the WD, which makes them to mostly emit in the B$_{435}$ band.
Since faint CVs should have less luminous donors, which, in turn, suggests lower donor masses (${\lesssim0.2-0.3}$~\Msun), they likely are below the orbital period gap, some might even be close to the orbital period minimum. 

\vspace{0.0cm}

On the other hand, in non-core-collapsed clusters ($\omega$~Cen and 47~Tuc), the separation between bright and faint CVs is not evident.
In these clusters, there seems to exist only one CV population, which is mostly composed of faint systems.
In fact, in the case of 47 Tuc, the CVs are such faint that most of them could not be detected even in the B$_{435}$ band. Instead, most systems have only been identified using NUV observations (Rivera Sandoval et al. 2018~\cite{Rivera_2018}).
Thus, if one wants to identify faint CVs, where the contribution of the companion is little, the best way is by doing a UV guided search. 
Note that, in Fig.~\ref{FigCMDobs}, we have plotted a faint and a bright population for the non-core-collapsed clusters.
This separation is artificial, not "intrinsic" like in the core-collapsed clusters. 
For $\omega$~Cen, we adopted a similar cut-off in the R$_{625}$ band, because the currently available data for this cluster was obtained with the same filters used in the studies of the core-collapsed clusters.
However, regarding 47~Tuc, as mentioned above, most CVs in this cluster are faint and detected for the first time in UV filters (22 systems).
Therefore, instead of choosing a cut-off in the R$_{625}$ band (as in the other clusters), the cut-off was made on their reddest NUV band, i.e. ${B_{390}\simeq9}$~mag.

\vspace{0.0cm}
\begin{figure}[t]
  \begin{center}
    \includegraphics[width=0.45\linewidth]{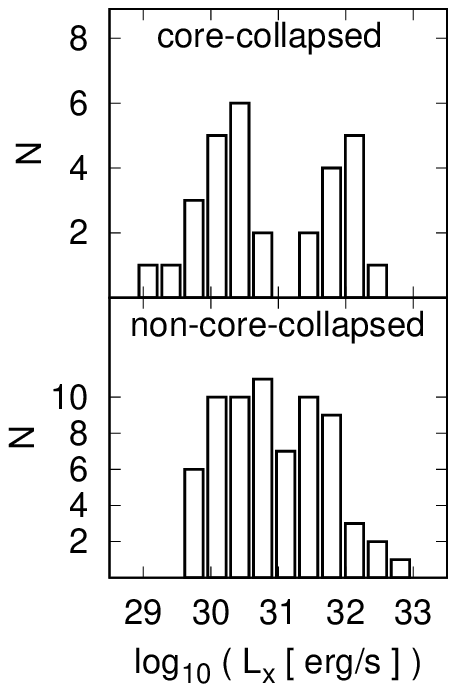}
    \includegraphics[width=0.45\linewidth]{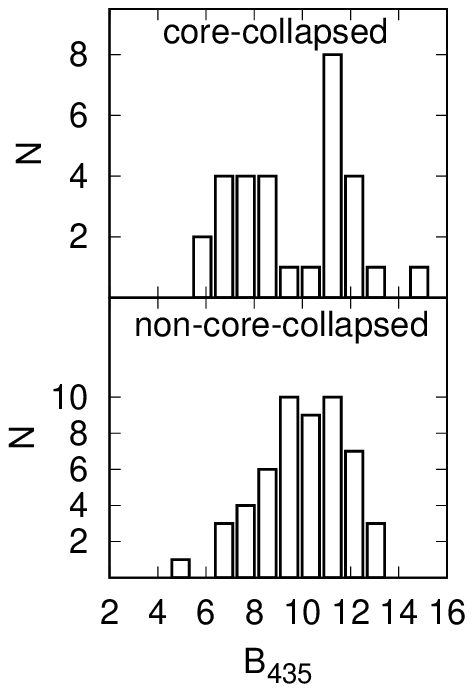}
    \end{center}
 \caption{Distribution of X-ray luminosities ($L_X$), in the [0.5-6.0]~keV band, and absolute $B_{435}$ magnitudes for the CVs in NGC~6397 (Cohn et al. 2010~\cite{Cohn_2010}), NGC~6752 (Lugger et al. 2017~\cite{Lugger_2017}), $\omega$~Cen (Cool et al. 2013~\cite{Cool_2013}), and 47~Tuc (Rivera Sandoval et al. 2018~\cite{Rivera_2018}). Absolute magnitudes are the same as those in Fig.~\ref{FigCMDobs} and X-ray luminosities are from the above mentioned papers and references therein. Core-collapsed GCs exhibit a clear bimodal distributions for the two properties. Non-core-collapsed GCs, on the other hand, have a bimodal distribution only for $L_X$, with no conspicuous peaks, and a unimodal distribution for $B_{435}$.
For the X-ray distributions just considered the systems for which a counterpart has been identified or at least there is a plausible candidate.
 }
 \label{FigLxRBobs}
\end{figure}
\vspace{0.0cm}

With respect to the X-ray luminosity, interestingly, we do not see a drastic difference between core-collapse and non-core-collapsed GCs (left-hand panel of Fig.~\ref{FigLxRBobs}).
Indeed, NGC~6397, NGC~6752, 47~Tuc and $\omega$~Cen present a bimodal distribution, i.e. a population of faint X-ray CVs and another population of bright X-ray CVs, even though the distribution of non-core-collapsed GCs is characterized by no conspicuous peaks.
It is important to mention that $\omega$~Cen has a large influence in the panel for non-core-collapsed GCs, as the counterparts for many faint X-ray sources have not been identified yet, but many of which are likely CVs. Adding those systems will likely make more evident the X-ray bimodality of non-core-collapsed GCs.
See Rivera Sandoval et al. (2018 \cite{Rivera_2018}) for a comparison between the other 3 clusters, where the X-ray bimodality is much more evident.

This bimodal X-ray distribution could be explained by mass transfer rate changes during the CV evolution and by the nature of the CVs themselves.
As a CV evolves from long towards shorter orbital periods, its mass transfer rate is expected to continuously drop.
Since the X-ray flux in CVs is intrinsically connected with the amount of matter accreted by the WD (e.g. Patterson \& Raymond 1985~\cite{Patterson_1985}; Hayashi \& Ishida 2014~\cite{Hayashi_2014}) and, in turn, with the mass transfer rate, the X-ray luminosity is also expected to drop as a CV evolves.
In this way, optically bright CVs would be more X-ray luminous, and most optically faint CVs would be less X-ray luminous, in both, core-collapsed and in non-core-collapsed GCs.
We also notice that some CVs in the four GCs have also very high X-ray luminosity ($L_X\gtrsim10^{32}$~erg~s$^{-1}$) and might be magnetic CVs.
Since magnetic CVs are brighter than DNe in X-rays (see, e.g. Mukai 2017~\cite{Mukai_2017} for a review on X-ray properties of accreting WDs), intermediate polars being the brightest amongst all CVs, it seems reasonable to expect that at least a part of the bright CVs in these four GCs would be composed of magnetic CVs, instead of only DNe.

Regarding the optical magnitudes from the right-hand panel of Fig.~\ref{FigLxRBobs}, 
we can see that in core-collapsed GCs, the luminosity distribution in $B_{435}$ is bimodal, which is consistent with the fact that in those GCs, two populations are observed in the CMD.
On the other hand, that distribution in non-core-collapsed GCs is unimodal and there is a deficit of bright CVs $(B_{435}\lesssim9)$, which indicates that core-collapsed clusters have relatively more bright CVs than non-core-collapsed GCs. We remind the reader that most CVs in 47 Tuc have been identified in bluer bands, that is why the bimodal distribution does not look as evident when using $B_{435}$. 
However, we have made a comparison in such a filter because is the middle-ground for both types of clusters considering the bands in which they have been studied in the literature. 
%

\subsubsection{Spatial distribution}
\label{Sec2.2.2}

\begin{figure}[t]
  \begin{center}
    \includegraphics
    [width=0.99\linewidth]
    {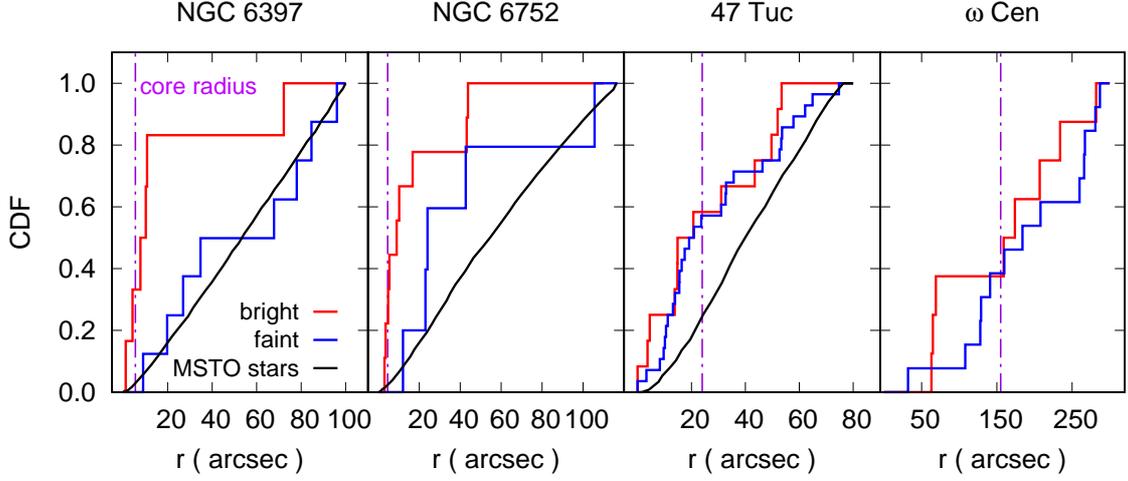}
    \end{center}
 \caption{Cumulative radial distributions for selected stellar groups in core-collapsed GCs (NGC~6397 and NGC~6752) and non core-collapsed GCs ($\omega$~Cen and 47~Tuc). Figure produced with data from Cohn et al. (2010~\cite{Cohn_2010}; NGC~6397), Lugger et al. (2017~\cite{Lugger_2017}; NGC~6752), Cool et al. (2013~\cite{Cool_2013}, Henleywillis et al. (2018 \cite{Henleywillis_2018}); $\omega$~Cen), and Rivera Sandoval et al. (2018~\cite{Rivera_2018}; 47~Tuc). The vertical magenta lines indicate each cluster core radius, the black lines correspond to MS stars close to the turn-off point (MSTO), the red lines to the bright CVs and the blue lines to the faint CVs. For $\omega$~Cen, information about MSTO is not available. Notice that in NGC~6397, bright CVs are more centrally concentrated than faint CVs, which are as concentrated as MSTO stars. In NGC~6752, bright CVs are more centrally concentrated than faint CVs, which, in turn, are more centrally concentrated than MSTO stars. In $\omega$~Cen, bright CVs are slightly more centrally concentrated than faint CVs, the difference between the two populations being larger inside the core. In 47~Tuc, on the other hand, bright and faint CVs are similarly segregated.}
 \label{FigPosobs}
\end{figure}

Another interesting property associated with bright and faint CVs in GCs is related to their spatial distributions (Fig.~\ref{FigPosobs}), for the GCs NGC~6397, NGC~6752, 47~Tuc and $\omega$~Cen. With the exception of $\omega$~Cen, distribution of \msto~are also shown.
We can also see in this figure the differences between the CV populations in these four GCs, with respect to the radial distribution of CVs.
Starting with the differences between bright and faint CVs in the core-collapsed GCs NGC~6397 and NGC~6752, bright CVs are more centrally concentrated than faint CVs. In NGC~6397 faint CVs have similar radial distribution to MS stars close to the turn-off point. In NGC~6752 faint and bright CVs are more centrally concentrated than MS stars.
Regarding the non-core-collapsed GCs, all CVs in 47 Tuc are equally concentrated. All of them actually more concentrated than MS stars close to the turn-off point.
In the case of $\omega$~Cen, when we consider all the CVs identified in the cluster (Cool et al. 2013~\cite{Cool_2013}; Henleywillis et al. 2018~\cite{Henleywillis_2018}), we see that bright CVs are slightly more concentrated than faint CVs in the core of the cluster. Unfortunately the information of the spatial distribution of the MS stars is not easily available in the literature. 

The different spatial distributions of faint and bright CVs in the four GCs considered here are most likely explained by the different average masses of each population in each cluster and the cluster relaxation times
(proxy for the GC dynamical age), which is consistent with recent numerical simulations (e.g. Hong et al. 2017~\cite{Hong_2017}; Belloni et al. 2019~\cite{Belloni_2019}).
In Table~\ref{TableCVMasses}, we summarise the cluster core/half-light relaxation times and the results from maximum-likelihood fitting of generalised King models to the surface density profiles available for NGC~6397, NGC~6752 and 47~Tuc. Note that the core-collapsed globular clusters have shorter relaxation times and have a substantial difference in the inferred mass of their two CV populations. An explanation about the differences between radial distributions will be discussed in Section~\ref{sec:belloni2019}.

\vspace{0.0cm}
\vspace{0.0cm}

\vspace{0.5cm}

\begingroup

\setlength{\tabcolsep}{5.0pt} 
\renewcommand{\arraystretch}{1.2} 

\begin{table}[t]
\centering
\caption{Maximum-likelihood fitting of generalized King models to the surface density profiles performed by Cohn et al. (2010~\cite{Cohn_2010}; NGC~6397),  Lugger et al. (2017~\cite{Lugger_2017}; NGC~6752) and Rivera Sandoval et al. (2018~\cite{Rivera_2018}; 47~Tuc). Regarding the cluster properties, we show in the first columns the cluster core ($T_{\rm core}$) and half-light ($T_{\rm half{\text -}light}$) relaxation times, as provided in the Harris (1996~\cite{Harris_1996}) catalogue, where we also included $\omega$~Cen for comparison. Concerning the results from the fitting scheme, we show in the remaining columns the radius within which the stellar groups have been considered, the number of systems in each stellar group involved in the fitting scheme, the mass ratio with respect to the MS stars close to the turn-off point (MSTO), and the average mass of each stellar group.}
\label{TableCVMasses}
\begin{tabular}{cccccccc}
\hline
%
%
\multirow{2}{*}{GC}  & $T_{\rm core}$ & $T_{\rm half{\text -}light}$  & radius    & \multirow{2}{*}{stellar group} & \multirow{2}{*}{number} & \multirow{2}{*}{mass ratio} & average mass  \\
    & (Myr)              &        (Myr)              & (arcsec)   &              &       &             &  (\Msun)   \\
\hline
\multirow{3}{*}{NGC~6397} & \multirow{3}{*}{0.09} & \multirow{3}{*}{400} & \multirow{3}{*}{100} & 
MSTO       & $1111$ & $1.00$        & $0.80\pm0.05$ \\
                          &                       &                      &                      & 
faint CVs  &    $8$ & $1.04\pm0.24$ & $0.83\pm0.20$ \\
                          &                       &                      &                      &
bright CVs &    $6$ & $1.83\pm0.26$ & $1.46\pm0.22$ \\
\hline
\multirow{3}{*}{NGC~6752} & \multirow{3}{*}{7} & \multirow{3}{*}{740} & \multirow{3}{*}{115} & 
MSTO       & $10016$ & $1.00$        & $0.80\pm0.05$ \\
                          &                       &                      &                      &
faint CVs  &     $5$ & $1.27\pm0.24$ & $1.02\pm0.19$ \\
                          &                       &                      &                      &
bright CVs &     $9$ & $2.03\pm0.35$ & $1.62\pm0.28$ \\
\hline
\multirow{3}{*}{47~Tuc}   & \multirow{3}{*}{70} & \multirow{3}{*}{3\,550} & \multirow{3}{*}{80}  & 
MSTO       & $34358$ & $1.00$        & $0.85\pm0.05$ \\
                          &                       &                      &                      &
faint CVs  &    $25$ & $1.60\pm0.24$ & $1.36\pm0.20$ \\
                          &                       &                      &                      &
bright CVs &    $11$ & $1.70\pm0.36$ & $1.45\pm0.31$ \\
\hline
$\omega$~Cen  & 3\,980 & 12\,300 & --- & --- & --- & ---  & --- \\
\hline
\end{tabular}
\end{table}

\endgroup

\vspace{0.5cm}

\subsection{Correlation between the number of cataclysmic variables and the cluster stellar encounter rate}
\label{Sec2.3}

A large fraction of `exotic' sources in GCs are X-ray emitters (see Heinke et al. 2020 \cite{Heinke_2020} and references therein, and also van den Berg 2020~\cite{vandenBerg_2020}, for a recent review on X-ray sources in Galactic GCs and old open clusters) and they can be separated between {\it bright X-ray sources} $(L_X\gtrsim10^{36}$ erg s$^{-1})$ and {\it faint X-ray sources} $(L_X\lesssim10^{34}$ erg s$^{-1})$.
The population of bright X-ray sources are dominated by low-mass X-ray binaries, which are either black holes or neutron stars accreting 
from MS stars at very high rates (e.g. Tauris \& van den Heuvel 2006~\cite{Tauris_2006}) and (perhaps in a less amount) ultra-compact X-ray binaries, in which either a neutron star or a black hole accretes from a Roche-lobe filling WD (e.g. Bildsten \& Deloye 2004~\cite{Bildsten_2004}; Cartwright et al. 2013~\cite{Cartwright_2013}; Heinke et al. 2013~\cite{Heinke_2013}).
Regarding faint X-ray sources, it was proposed since the 1980s and 1990s that most of them should be either CVs (e.g. Hertz \& Grindlay 1983a~\cite{Hertz_1983a}) or chromospherically active binaries (ABs; e.g. Bailyn, Grindlay \& Garcia 1990~\cite{Bailyn_1990}), which are binary stars with spectral types later than F characterized by a strong chromosphere, transition region, and coronal activity (e.g. Eker et al. 2008~\cite{Eker_2008}).
In addition, to account for the most luminous objects among faint X-ray sources, quiescent low-mass X-ray binaries (e.g. Heinke et al. 2003 \cite{Heinke_2003}; Verbunt \& Lewin 2006 \cite{2006verbunt}) and millisecond pulsars (e.g. Grindlay \& Bogdanov, 2009 \cite{Grindlay2009}) have been proposed.
Only with \chandra~and \hst~the uncertainty related to the nature of faint X-ray sources has been significantly reduced, most of them being indeed CVs and ABs (e.g. Cheng et al. 2018a~\cite{Cheng_2018}, and references therein for a recent X-ray study).
We would like to mention that, apart from these sub-populations comprising faint X-ray sources, in many cases there is contamination from foreground/background objects (e.g. Heinke et al. 2020~\cite{Heinke_2020}, and references therein), which are physically unrelated to the GCs and which X-ray emissivity can erroneously be accounted as part of the GC.

A way of addressing whether X-ray sources are potentially created in GCs through dynamical interactions is by means of the \emph{stellar encounter rate} $\Gamma$.
It depends on the central density, the core radius and the central velocity dispersion, which makes it a better way of quantifying the rate of dynamical interactions in a particular cluster (e.g. Verbunt \& Hut 1987~\cite{Verbunt_1987}, Pooley et al. 2003~\cite{Pooley_2003}, Davies, Piotto \& De Angeli 2004~\cite{Davies_2004}) than individual properties such as the core radius, the concentration or the central density.
If dynamical interactions are supposed to play any significant role in the formation of X-ray sources, then one could expect a correlation between the number of sources within a cluster and $\Gamma$.
However, this interpretation should be taken with a grain of salt, because $\Gamma$ is usually not constant during the cluster evolution and dynamics not always contribute in creating binaries.
Indeed, $\Gamma$ oscillates due to several episodes of core collapse during the cluster evolution (e.g. Hong et al. 2017~\cite{Hong_2017}; Beccari et al. 2019~\cite{Beccari_2019}), which makes their present-day values not necessarily comparable to any of its previous values during the cluster evolution over its life-time (most GCs with ages between $\sim11-13$~Gyr; van den Berg et al. 2013~\cite{VandenBerg_2013}).
Moreover, dynamics likely not only play a role in creating X-ray sources over the cluster life-time (especially close to the present day), but also in destroying their progenitors from primordial binaries (e.g. Davies 1997~\cite{Davies_1997}; Milone et al. 2012~\cite{Milone_2012}; de Grijs et al. 2013\cite{Grijs_2013}; Leigh et al. 2015~\cite{Leigh_2015}; Cheng et al. 2018a~\cite{Cheng_2018}; Belloni et al. 2019~\cite{Belloni_2019}).
Therefore, the interplay between production of X-ray sources and destruction of their primordial progenitors over the cluster life-time is not easily addressed with only present-day GC properties.

Several works attempted to verify whether a correlation between the number of X-ray sources and the $\Gamma$ does exist.
Verbunt \& Hut (1987~\cite{Verbunt_1987}) showed that bright X-ray sources in GCs seem consistent with being formed through dynamical interactions.
There are currently 21 bright X-ray sources in 15 GCs, being eight persistent and 13 transients (van den Berg 2020~\cite{vandenBerg_2020}), and they are still consistent with being dynamically formed, especially considering their mass density, which is much higher than in the Galactic disc.
Pooley et al. (2003~\cite{Pooley_2003}) analysed \chandra~observations of 12 GCs and found that the number of X-ray sources correlates with $\Gamma$.
These authors concluded that the X-ray populations in GCs are largely dynamically formed.
Maxwell et al. (2012~\cite{Maxwell_2012}) updated the values of $\Gamma$
and the numbers detected in more recent studies with respect to Pooley et al. (2003~\cite{Pooley_2003}), but found similar trends when considering the same 12 GCs.

Bahramian et al. (2013~\cite{Bahramian_2013}) re-derived the values of $\Gamma$ for 124 GCs by deprojecting cluster surface brightness profiles to estimate luminosity density profiles, which allowed them to treat equally core-collapsed and non-core-collapsed GCs.
By comparing their results on $\Gamma$ to the numbers of X-ray binaries in GCs, these authors suggested that X-ray binaries are relatively under-produced in core-collapsed clusters, which indicates that dynamical interactions play an important role in destroying these binary progenitors compared to non-core-collapsed clusters.
This was also investigated by Lugger et al. (2017~\cite{Lugger_2017}), who analysed the X-ray sources abundance in NGC~6752 and compared it to other 9 GCs. These authors showed that indeed core-collapsed clusters seem to have fewer X-ray sources relative to their computed encounter rates.
Previous to these works, and considering a larger sample of 63 GCs, Pooley (2010~\cite{Pooley_2010}) concluded that core-collapsed clusters have more sources for a given $\Gamma$, unlike what was found by
Bahramian et al. (2013~\cite{Bahramian_2013}) and Lugger et al. (2017~\cite{Lugger_2017}).
However, that study was not fully conclusive given that a test of the robustness of the author's findings was not performed. 

Regarding faint X-ray sources, Pooley \& Hut (2006~\cite{Pooley_2006}) analysed 23 GCs and found that a population based on only primordial binaries is ruled out and a population based on primordial plus dynamical binaries more likely represents the data. In particular, they found correlation between the mass-normalized number of faint X-ray sources and the mass-normalized stellar encounter rate.
This picture has been recently challenged by Cheng et al. (2018a~\cite{Cheng_2018}), who investigated the X-ray emissivity (total X-ray luminosity per unit mass) from faint X-ray sources in GCs with \emph{Chandra} archival data.
Unlike Pooley \& Hut (2006~\cite{Pooley_2006}), who focused on the individually resolved sources, their approach is similar to those of Verbunt (2001~\cite{Verbunt_2001}) and Ge et al. (2015~\cite{Ge_2015}), and assumes that the X-ray emissivity is a reasonable proxy of the source abundance.
These authors analysed 69 GCs, which correspond to nearly half of the known Galactic GC population, and to a sample $\sim6$ and $\sim3$ times larger than those investigated by Pooley et al. (2003~\cite{Pooley_2003}) and Pooley \& Hut (2006~\cite{Pooley_2006}), respectively.
Chen et al., found that there is not a significant correlation between the faint X-ray source abundance and the mass-normalized stellar encounter rate, which clearly disagrees with Pooley \& Hut (2006~\cite{Pooley_2006}).
In this way, Cheng et al. (2018a~\cite{Cheng_2018}) have shown that dynamical interactions are less dominant than previously believed, and that the primordial formation channel has a substantial contribution. 
That result was also supported more recently by Heinke et al. (2020~\cite{Heinke_2020}), though these authors also discuss some caveats in the work by Cheng et al. (2018a~\cite{Cheng_2018}) which could have influenced their conclusions. 
Another interesting and striking result by Cheng et al. (2018a~\cite{Cheng_2018}) is that, unlike what has been previously thought, the faint X-ray populations are under-abundant in GCs with respect to the Solar neighbourhood and Local Group dwarf elliptical galaxies.
This indicates that dynamical destruction of faint X-ray source progenitors is a non-negligible effect in GCs, although the net balance between formation and destruction through dynamical interaction is not easily addressed observationally.
%

\section{What do we know from theory and simulations?}
\label{SecSIM}

The first efforts to determine the properties of CVs in GCs usually focused on particular formation channels and did not had GC evolution (e.g. Bailyn, Grindlay \& Garcia 1990~\cite{Bailyn_1990}; Stefano \&  Rappaport 1994~\cite{Stefano_1994}; Davies 1995~\cite{Davies_1995}; Davies \& Benz 1995~\cite{Davies_1995b}; Davies 1997~\cite{Davies_1997}; Ivanova et al. 2006 \cite{Ivanova_2006}).
Following those, Shara \& Hurley (2006 \cite{Shara_2006}) performed the first realistic GC simulations with focus on CVs.
For a summary of these works in a comparative fashion, we recommend the review by Benacquista \& Downing (2013~\cite{Benacquista_2013}).
More recently, CVs in GCs have been investigated in GC simulations performed with the \mocca~code by Hong et al. (2017 \cite{Hong_2017}) and Belloni et al. (2016 \cite{Belloni_2016a}; 2017a \cite{Belloni_2017a}; 2017b \cite{Belloni_2017b}; 2019 \cite{Belloni_2019}), which we will discuss in more details in what follows.


The \mocca~code
is based on the orbit-averaged Monte Carlo technique
for star cluster evolution developed by H\'enon (1971~\cite{Henon_1971}), which was further improved by Stod{\'o}{\l}kiewicz (1982~\cite{Stodolkiewicz_1982}, 1986~\cite{Stodolkiewicz_1986}), and then developed even further by Giersz et al. (1998~\cite{Giersz_1998}, 2001~\cite{Giersz_2001}, 2006~\cite{Giersz_2006}, 2008~\cite{Giersz_2008}, 2013a~\cite{Hypki_2013}, 2013b~\cite{Giersz_2013}).
The current version includes the {\sc fewbody} code (Fregeau et al. 2004~\cite{Fregeau_2004}) to perform numerical scattering experiments of small-number gravitational interactions and the \bse~code  (Hurley et al. 2000~\cite{Hurley_2000}, 2002~\cite{Hurley_2002}) to evolve stars and binaries. 
\mocca~assumes a point-mass Galactic potential with total mass equal to the enclosed Galaxy mass inside a circular orbit at the specified Galactocentric radius, and uses the description of escape processes in tidally limited clusters which follows the procedure derived by Fukushige \& Heggie (2000~\cite{Fukushige_2000}). 
Due to its speed, efficiency and accurate coverage of the relevant parts of the parameter space, \mocca~is ideal for performing big surveys aimed at modelling large populations of CVs in many GCs, and for studying in detail the influence of the host cluster environment on their properties.

Before proceeding further, it is convenient to dedicate a few words to the concept of {\it initial binary population}, since it is a rather important concept in GC evolution modelling and was quite explored by Belloni et al. (2016~\cite{Belloni_2016a}, 2017a~\cite{Belloni_2017a}, 2017b~\cite{Belloni_2017b}, 2019~\cite{Belloni_2019}) and Hong et al. (2017~\cite{Hong_2017}) in their simulations.
The initial binaries in GCs follow determined distributions for their parameters: semi-major axis, eccentricity, stellar masses, mass ratio, and orbital period.
Hereafter, the \ibp~is the set that contains all initial binaries, in a given initial cluster, associated with specific distributions for their parameters. 
One of the aims of these investigations was to test whether a particular \ibp~ would be a better input for star cluster simulations, or if they would provide comparable results.
The \textit{Kroupa} \ibp~was derived by Kroupa (1995a~\cite{Kroupa_1995a}, 1995b~\cite{Kroupa_1995b}, 1995c~\cite{Kroupa_1995c}, 2008~\cite{Kroupa_INITIAL}, 2013~\cite{Kroupa_2013}) and has the following characteristics: (i) the orbital period distribution monotonically increases towards $\sim10^8$~days; (ii) short-period ($\lesssim10^3$~days) binaries tend to have low eccentricity and long-period ($\gtrsim10^3$~days) binaries to be thermalized, and (iii) the mass ratio distribution is roughly flat with a huge peak close to $\approx1$ (see also Marks, Kroupa \& Oh 2011~\cite{Marks_2011}, for more details).
On the other hand, the \textit{Standard} \ibp~corresponds to 'flat' distributions typically adopted in star clusters simulations: (i) a uniform distribution for the mass ratio, (ii) a log-uniform distribution for the semi-major axis, and (iii) a thermal distribution for the eccentricity.
While models set with the Kroupa \ibp~should have nearly $100\%$ primordial binaries, by construction, those set with the Standard \ibp~usually have low binary fractions ($\lesssim10\%$) in typical star cluster simulations.

\subsection{Simulations by Belloni et al. (2016, 2017a, 2017b)}

Belloni et al. (2016~\cite{Belloni_2016a}, 2017a~\cite{Belloni_2017a}, 2017b~\cite{Belloni_2017b}) investigated 12 GC models simulated with the \mocca~code. 
These models differ mainly with respect to the initial concentration (sparse, dense, very dense), initial binary population (`Kroupa' and `Standard'), and the common-envelope (CE) parameters (`very high' and `high', where `very high' means $\alpha_{\rm CE}=3$ and $\alpha_{\rm rec}=0.5$ and `high' means $\alpha_{\rm CE}=1$ and $\alpha_{\rm rec}=0$).
Here, $\alpha_{\rm CE}$ and $\alpha_{\rm rec}$ are the CE efficiency and fraction of recombination energy, respectively.
Regarding their goals, we can summarize the main issues they tried to address in the following questions: (i) {\it Why does modelling predict hundreds of DNe in GCs, but only a few eruptions have been observed?} (ii) {\it Are GC CVs predominantly magnetic?} (iii) {\it What are the formation channels leading to bright and faint CVs?} (iv) {\it Can CVs indeed help in constraining initial star cluster conditions and the \ibp?} (v) {\it How dynamics affect CV progenitors?} (vi) {\it Are GC CVs similar to Galactic-field CVs?}

%
%

With respect to points (i) and (ii), Belloni et al. (2016~\cite{Belloni_2016a}) showed that the fact that there is an apparent paucity of DNe in GCs should actually be translated as a deficit of DN outbursts, not DNe necessarily.
This would suggest that if the majority of GC CVs are DNe, then their duty cycles should be extremely short and, in turn, extremely faint.
This is indeed predicted in their simulations, where the majority of GC CVs are period bouncers and that GC CV duty cycles are extremely low, which implies that the probability of detecting GC CVs during outburst is, in turn, quite low.
This suggests that most GC CVs are DNe, not magnetic CVs, which indicates that there is no need to claim for an overabundance of magnetic CVs in GCs, in comparison with the Milky Way field.
In this way, these authors showed that the null results with respect to the detection of DNe via their variabilities through outbursts is actually predicted and does not correspond to a problem between theory and observations.
Going further, as claimed before while discussing the observational properties of GC CVs, the results from the numerical simulations also support the scenario in which searching for CVs by means of the multi-wavelength approach is by far more efficient than trying to identify DNe via their variability through outbursts.
This is because it is predicted that only a fraction of them would have duty cycles long enough to be identified as erupting DNe and this type of approach has already been shown to be rather unsuccessful (Shara et al. 1996~\cite{Shara_1996}; Pietrukowicz et al. 2008~\cite{Pietrukowicz_2008}; Shara [private communication]) given the cadences and filters used so far.
In any event, a final and required step in this regard is to try to confirm spectroscopically those several candidates revealed with the multi-wavelength approach.

%
%

Regarding the point (iii), Belloni et al. (2017b~\cite{Belloni_2017b}) found from their simulations that, by assuming the classical consequential angular momentum loss
(King \& Kolb 1995~\cite{King_1995}) and high CE efficiency, bright CVs in GCs are young and mainly formed due to exchanges.
Faint CVs are a mix of CVs formed through different formation channels (i.e. CE evolution, weak and strong dynamical interactions), which cannot easily be disentangled due to the average CV evolution time-scale (a few Gyr).
This fast evolution naturally erases the signature of the CV formation channel, which is not the case regarding bright CVs, since they did not have enough time to lose the characteristics from their formation channel. 
However, as argued for the WD masses obtained by Ivanova et al. (2006~\cite{Ivanova_2006}), this conclusion is also strongly dependent on the binary evolution assumption, especially the assumed angular momentum loss and stability criteria for dynamical mass transfer at the onset of the CV phase (e.g. Schreiber, Zorotovic \& Wijnen 2016~\cite{Schreiber_2016}).
Additionally, intermediate polars formed through binary evolution might have similar X-rays and photometric properties to dynamically formed CVs.
Thus, it is not always clear what is the formation channel from observations.

%
%

The point (iv) from the list is the most important one with respect to GC modelling, since it would allow us to use observational properties of CVs (and potentially other types of exotic binaries) to constraint initial star cluster conditions, including their \ibp{}s.
In the specific case of GCs, if the attempt of modelling particular clusters, say NGC~6397 and  47~Tuc, leads to several best-fitting models while looking only at dynamical properties, one could use the information related to CVs, provided from observation, in order to filter out even more the set of best-fitting models.
In general, if  one fails to reproduce the CV properties within the best-fitting dynamical models, variations of the \ibp, including the initial mass function, and/or binary evolution parameters might improve the models and in turn our understanding with respect to these two extremely important aspects of GC modelling.
This point could only be addressed in more details in Belloni et al. (2019~\cite{Belloni_2019}), whose results will be discussed later in this section.

%
%

Moving forward, on point (v), Belloni et al. (2017b~\cite{Belloni_2017b}) found that {\it the denser the cluster is initially, the smaller the number of CVs formed through binary evolution alone}.
In fact, for all their clusters, {\it any reduction in the relative numbers of CV progenitors is correlated with the initial cluster density}, which is associated with the role of dynamical interactions in destroying CV progenitors, which is, in turn, related to the cluster soft-hard boundary.
This boundary is set when the average binary binding energy equals to the average cluster kinetic energy.
This separation is thus intrinsically related to the interplay between the binary binding energies with respect to host GC properties.
Pragmatically, it corresponds to the orbital period separating hard and soft binaries (Heggie \& Hut 2003~\cite{HeggieBOOK}).
\textit{Hard} binaries are very strongly bound and are not expected to go through disruptive encounters, while \textit{soft} binaries, on the other hand, are very weakly bound and tend to be destroyed in dynamical interactions.
Some binaries have orbital periods comparable to the hard-soft boundary and can sometimes be destroyed or significantly altered.
Most binaries evolve according to the Heggie--Hills law: {\it hard binaries get harder, while soft binaries get softer, after dynamical interactions} (Heggie 1975~\cite{Heggie_1975}, Hills 1975~\cite{Hills_1975}), which implies that soft binaries tend to be eventually disrupted at some point.
This law has been recently received support from observational studies of faint X-ray sources in GCs (Cheng et al. 2018b~\cite{Cheng_2018b}).

%
%

Concerning the point (vi), Belloni et al. (2017a~\cite{Belloni_2017a}) found that comparing GC CVs with Galactic-field CVs might provide misleading results, as GC CVs should be different from Galactic field CVs.
Indeed, dynamics can extend the parameter space applicable to CV progenitors and allow binaries that would not otherwise become CVs to evolve into CVs, which then could make at least some GC CVs intrinsically different from most Galactic-field CVs.
In addition, in very dense environments, many primordial binaries that are CV progenitors are likely to be destroyed due to dynamical interactions or have their formation either accelerated or retarded due to dynamics.
This characteristic of having GC CVs different from Galactic-field CVs holds for all GC environments, including sparse GCs dominated by non-dynamical CVs.
Even though sparse clusters have more CVs formed through CE evolution and less destroyed CV progenitors, relative to denser clusters, one would still have a different population with respect to the Galactic field population due to different age and star formation histories in both environments.
While in GCs there is an initial burst (or several bursts at the beginning separated by short time-scales), in the Milky Way disc, star formation takes place more or less continuously throughout the disc evolution (e.g. Weidner, Kroupa \& Larsen 2004~\cite{Weidner_2004}, Recchi \& Kroupa 2015~\cite{Recchi_2015}, Schulz, Pflamm-Altenburg \& Kroupa 2015~\cite{Schulz_2015}).
Additionally, measured ages of GCs cluster around $\sim12$~Gyr (van den Berg et al. 2013~\cite{VandenBerg_2013}), while the Milky Way disc age is measured to be $\sim10$~Gyr (Kilic et al. 2017~\cite{Kilic_2017}).
These two different aspects make GC CVs intrinsically older than
Milky Way disc CVs.

\subsection{Simulations by Hong et al. (2017)}

Hong et al. (2017~\cite{Hong_2017}) investigated the influence of GC properties on CV dynamical formation and CV radial distribution with the \mocca~code.
Their main goal was actually to investigate the apparently observed correlation between the number of CVs and the stellar encounter rate (Pooley \& Hut 2006~\cite{Pooley_2006}).
These authors analysed 81 GC models evolved with the \mocca~code assuming the Kroupa \ibp, and a very high CE efficiency, i.e. $\alpha_{\rm CE}=3$ and $\alpha_{\rm rec}=0.5$.
Their initial GC model vary with respect to masses, half-mass radii, Galactocentric distances, and primordial binary fractions.
This is the first investigation of GC CVs in which a considerably large amount of realistic GC models have been studied, even though it is still limited with respect to binary evolution models/assumptions, as we shall see.

We should start discussing their results by addressing the implications of the assumed CE efficiency and \ibp.
These authors found that \textit{all} CVs in their models are dynamically formed.
This is consistent with Belloni et al. (2016~\cite{Belloni_2016a}) who showed that very high CE efficiency coupled with the Kroupa \ibp~does not produce CVs without dynamics.
The reason is a combination of two things.
First, the orbital period distribution in the Kroupa \ibp~is a monotonically increasing function (e.g. Marks \& Kroupa 2011~\cite{Marks_2011b}).
This implies that the majority of the binaries in the Kroupa \ibp~have orbital periods longer than $10^3$~days ($\approx83\%$).
Second, high CE efficiency leads to longer orbital periods after the CE evolution. 
This causes post-CE binaries, in general, to remain in the detached phase for longer time-scales.
Regarding CV progenitors, in particular, the pre-CV life-time is extended, being longer, for higher CE efficiencies .
This causes binaries in the Kroupa \ibp~to become CVs \textit{only} at a time-scale longer than the Hubble time.

Due to the large amount of GC models analysed, Hong et al. (2017~\cite{Hong_2017}) found two important statistical properties: (i) dynamical exchange is the mechanism responsible for the correlation between the stellar encounter rate and the number of CVs found by Pooley \& Hut (2006~\cite{Pooley_2006}), and (ii) bright CVs are more centrally concentrated than \msto, in agreement with observations (Cohn et al. 2010~\cite{Cohn_2010}, Lugger et al. 2017~\cite{Lugger_2017}, Rivera Sandoval et al. 2018~\cite{Rivera_2018}).

Concerning point (i), the stellar encounter rate is initially large (since the GC models are initially compact) and decreases during the GC early evolution (cluster expansion occurs because of stellar evolution, i.e. mass loss). 
As the GC keeps evolving, the stellar encounter rate stays approximately constant and slightly increases again as the GC approach core collapse (see their fig.~1). 
Thus, when the stellar encounter rate substantially increases, the density in the central parts also considerably increases and, consequentially so does the number of dynamical interactions.
Therefore, it is expected a consequential increase in the number of dynamically formed CVs, for large stellar encounter rates.
In their fig.~3, Hong et al. (2017~\cite{Hong_2017}) showed the number of CVs versus the stellar encounter rate, both normalized with respect to the GC masses.
There is a clear correlation between the two quantities while considering only CVs formed via dynamical exchanges.
On the other hand, there is no correlation (or very weak, if at all) for CVs formed through a combination of several weak and/or a few strong dynamical interactions, followed by CE evolution.
These authors could then conclude that CVs formed via exchanges are those responsible for the observed increase in the mass-normalized number of CVs with the mass-normalized stellar encounter rate, answering the question made by Pooley \& Hut (2006~\cite{Pooley_2006}).
However, as we will discuss in Section~\ref{Sec4.3}, it is still not clear whether such a correlation is indeed real or just a result of small-number statistics. 
Additionally, Hong et al. (2017~\cite{Hong_2017}) did not take into account CVs formed without dynamics, which could potentially have led to different results for the whole CV population (top left-hand panel of their fig.~3).
This implies that it is also not clear whether such correlation can actually be modelled in numerical simulations, since the inclusion of primordial CVs could simply break the correlation found when only dynamical CVs are taken into account.

Regarding point (ii), in their fig.~4, Hong et al. (2017~\cite{Hong_2017}) presented the cumulative radial distribution of all present-day CVs in all their models, divided according to the donor mass and compared with \msto.
Their bright CVs (defined as those having donors heavier than $0.1$~\Msun) are, in general, more centrally concentrated than \msto.
They found that this feature is due to the CV masses (more massive than \msto) coupled to the remaining memory of the CV formation history and progenitor masses.
As we have seen in Section~\ref{Sec2.2}, this trend is consistent with the results obtained for NGC~6397 (Cohn et al. 2010~\cite{Cohn_2010}), NGC~6752 (Lugger et al. 2017~\cite{Lugger_2017}) and 47~Tuc (Rivera Sandoval et al. 2018~\cite{Rivera_2018}).
Interestingly, they also found that faint CVs are more centrally concentrated than \msto, which was not found for the core-collapsed GCs mentioned above, but found for 47 Tuc.
This then indicates that, despite the fact these authors analysed CV properties on a statistical basis, it still remained unclear the effect of mass segregation associated with CVs and CV progenitors and further investigations could address in more details the movement of CVs and CV progenitors during the GC evolution, for different GC and CV types.

\subsection{Problems in the simulations by Belloni et al. (2016, 2017a, 2017b) and Hong et al. (2017)}

The main weaknesses of the numerical simulations by Belloni et al. (2016~\cite{Belloni_2016a}, 2017a~\cite{Belloni_2017a}, 2017b~\cite{Belloni_2017b}) and Hong et al. (2017~\cite{Hong_2017}) were associated with the assumed initial GC conditions and adopted binary evolution parameters, mainly the \ibp, and the treatment of CV formation and evolution.

%
%
With respect to the initial GC conditions, these authors adopted the Kroupa \ibp~in their simulations, which can reproduce, to some extent, some observable properties of the binaries in different environments.
Nevertheless, before these works, the Kroupa \ibp~had never been tested against observed GC CMDs.
Askar et al. (2018~\cite{Askar_COCOA}) found that predicted present-day GC CMDs  \textit{always} exhibit an additional sequence caused by MS--MS twins, when the Kroupa \ibp~is adopted, which is not observed in all Galactic GCs (Milone et al. 2012~\cite{Milone_2012}).
Indeed, in GC numerical simulations with the Kroupa \ibp, synthetic CMD colour distributions exhibit a peak associated with binaries that have mass ratios $\approx1$. 
While the Kroupa \ibp~reproduces binary properties in star-forming regions, clusters and the Galactic field, the peak in the mass ratio distribution towards $\approx1$ found in GC simulations is not consistent with distributions derived from observations.
Additionally, Hong et al. (2017~\cite{Hong_2017}) have binary fractions smaller than $100\%$, which are not consistent within the framework in which this \ibp~was built.

%
%
With respect to CV formation, a rather high CE efficiency ($\alpha_{\rm CE}\geq1$) was adopted by these authors, besides assuming that half of the recombination energy ($\alpha_{\rm rec}=0.5$) helps in the envelope ejection.
There is significant evidence for $\alpha_{\rm CE}\sim0.2-0.3$ being a more realistic value regarding post-CE binaries harboring WDs, and there is usually no need to assume that the recombination energy should be used to assist expelling the CE, i.e. $\alpha_{\rm rec}\sim0$.
This set of parameters is consistent with recent investigations that have concluded that WD--MS binaries experience a strong orbital shrinkage during CE evolution (e.g. Zorotovic et al. 2010~\cite{Zorotovic_2010}, Toonen \& Nelemans 2013~\cite{Toonen_2013}, Camacho et al. 2014~\cite{Camacho_2014}, Cojocaru et al. 2017~\cite{Cojocaru_2017}).
Different choices of the CE parameters definitely change the number of CVs formed from primordial binaries and, in turn, GC CV properties as a whole, since dynamically formed CVs tend to have different properties from CVs formed primordially.

%
%
With respect to the treatment of CV evolution, the investigations by Belloni et al. (2016~\cite{Belloni_2016a}, 2017a~\cite{Belloni_2017a}) and Hong et al. (2017~\cite{Hong_2017}) were performed with the \bse~version having the caveats described in Belloni et al. (2017b~\cite{Belloni_2017b}, see their section~5.2).
Even though the original \bse~code is frequently used for binary population modelling of CVs and related objects (e.g. Meng \& Yang 2011~\cite{Meng_2011}, Zuo \& Li 2011~\cite{Zuo_2011}, Schreiber, Zorotovic \& Wijnen 2016~\cite{Schreiber_2016}, Zorotovic \& Schreiber 2017~\cite{Zorotovic_2017}), it is best suited to just model  the early phases of their evolution, i.e. from the zero-age MS until the formation of post-CE binaries (e.g. Chen et al. 2014~\cite{Chen_2014}, Zorotovic et al. 2016~\cite{Zorotovic_2016}). 
This is mainly because the \bse~code in its original form only includes simple prescriptions for the evolution of accreting compact-object systems and comprehensive testing of mass transfer phases in these systems was beyond the original scope. 
As such, fundamental ingredients of CV evolution were missing, and using that version of the \bse~code for CVs can easily lead to inaccurate predictions for, e.g. mass transfer rates, duty cycles, or the orbital period and donor mass distributions (Belloni et al. 2017b~\cite{Belloni_2017b}).

%
%
A final problem in the simulations by Belloni et al. (2016~\cite{Belloni_2016a}, 2017a~\cite{Belloni_2017a}, 2017b~\cite{Belloni_2017b}) is the amount of GC models.
Even though these authors were able to find interesting and important results, and even discuss about the nature of GC CVs, obtained from the analysis of a substantial amount of CVs, one would agree that 12 GC models are not representative and there is still a problem in this set of investigations with respect to small-number statistics involved in their GC modelling.

\subsection{Simulations by Belloni et al. (2019)}
\label{sec:belloni2019}

The fourth paper in the series by Belloni et al. intended to overcome all caveats previously discussed. 
Indeed, Belloni et al. (2019~\cite{Belloni_2019}) adopted the following elements: (i) consistent \ibp{}s, including the initial mass function preservation; (ii) a consistent CV evolution model; (iii) adequate choices for the CE parameters; (iv) a large set of GC models in order to improve statistical analyses and (v) a consistent variation of initial GC conditions.

%
%
Concerning point (i), these authors improved the {\sc mocca} code with respect to the Standard and Kroupa \ibp{}s.
In both cases, the initial mass function is now always preserved, applying a similar procedure as described in section~6.3 of Belloni et al. (2017c~\cite{Belloni_2017c}, see also Oh, Kroupa \& Pflamm-Altenburg 2015~\cite{Oh_2015} and Oh \& Kroupa 2016~\cite{Oh_2016}).
In addition, the Kroupa \ibp~was substantially improved (Belloni et al. 2017c~\cite{Belloni_2017c}, 2018a~\cite{Belloni_2018a}) in order to, not only solve the problem with respect to GC MS--MS binary distributions, but also provide a good agreement with respect to Galactic-field late-type MS binaries.

%
%
Regarding point (ii), the \bse~algorithm inside the \mocca~code was upgraded to allow accurate modelling of interacting binaries in which degenerate objects are accreting from low-mass MS donor stars, as described in Belloni et al. (2018b~\cite{Belloni_2018b}).
The main upgrades of the code correspond to a revised version of the mass transfer rate equation, of the radius increment/decrease of low-mass MS donors that is expected when mass transfer is turned on/off (and needed to explain the orbital period gap), new options for angular momentum losses, and different stability criteria for dynamical and thermal mass transfer from MS donors.
The \caml~prescription adopted was the one postulated by Schreiber, Zorotovic \& Wijnen (2016~\cite{Schreiber_2016}), which is currently the only model that can solve some long-standing CV problems, like the associated space density, the orbital period minimum, the orbital period distribution and the WD mass distribution.
Interestingly, it can also explain the existence of single helium-core WDs (Zorotovic \& Schreiber 2017~\cite{Zorotovic_2017}).
In addition, this updated version of the \mocca~code also includes updated prescriptions of massive star evolution as described in Giacobbo, Mapelli \& Spera (2018~\cite{Giacobbo_2018}).

%
%
Related to point (iii), these authors adopted three values for the CE efficiency (${\alpha_{\rm CE}=0.25}$, $0.5$ and $1.0$).
In addition, they assumed that none of the recombination energy helps in the CE ejection and that the binding energy parameter is determined based on the giant properties, as described in Claeys et al. (2014~\cite{Claeys_2014}, see their appendix~A).
Finally, regarding points (iv) and (v), Belloni et al. (2019~\cite{Belloni_2019}) simulated 288 GC models, whose initial configurations vary with respect to mass, density profile, Galactocentric distance, half-mass radius, \ibp, binary fraction, CE efficiency, and mass fall-back during compact object formation.
This is by far the largest sample of GC models ever analysed with respect to CVs.
Additionally, their present-day GC models cover a reasonable range of concentrations, central surface brightness and half-mass relaxation times (Belloni et al. 2019~\cite{Belloni_2019}, their fig.~1).
Therefore, such models are consistent with a substantial fraction of the real GCs and are, in turn, roughly representative of the Galactic GC population.

%
%
Regarding the main results from previous works discussed before, some were reproduced by Belloni et al. (2019~\cite{Belloni_2019}),
while others disagree.
For instance, unlike Ivanova et al. (2006~\cite{Ivanova_2006}), who found that the WD mass distribution for dynamical and non-dynamical CVs substantially differ (dynamical CV WDs being heavier than non-dynamical CV WDs), Belloni et al. (2019~\cite{Belloni_2019}) found that both CV types have similar WD mass distributions.
The reason for that is associated with the different stability criterion for dynamical mass transfer adopted in both simulations.
While in the former, fully conservative mass transfer was likely adopted, in the latter it was assumed that enhanced consequential angular momentum loss, as explained by Schreiber, Zorotovic \& Wijnen (2016~\cite{Schreiber_2016}), affects this criterion. 
In addition, Belloni et al. (2019~\cite{Belloni_2019}) confirmed previous findings by Belloni et al. (2016~\cite{Belloni_2016a}; 2017b~\cite{Belloni_2017b}) that most predicted CVs have very short DN outbursts in comparison with their recurrence time-scales, which implies very short duty cycles ($\lesssim6$~per~cent) and, in turn, extremely small probability of detection during outbursts.

%
%
With respect to the influence of dynamics, their simulations suggest that fewer CVs should be expected in dense GCs relative to the Milky Way field, due to the fact that destruction of CV progenitors is more important in GCs than dynamical formation of CVs. This is consistent with Shara \& Hurley (2006~\cite{Shara_2006}), who suggested a reduction of a factor of $2-3$, but the opposite of what was found by
Ivanova et al. (2006~\cite{Ivanova_2006}), who predicted an enhancement of a factor of $2-3$.
Belloni et al. (2019~\cite{Belloni_2019}) found a strong correlation, at a significant level, between the fraction of destroyed primordial CV progenitors and the initial GC stellar encounter rate, i.e. the greater the initial GC stellar encounter rate, the stronger the role of dynamical interactions in destroying primordial CV progenitors.
Moreover, the average mass density found by Belloni et al. (2019~\cite{Belloni_2019}) is ${\sim2\times10^{-5}}$~CVs~M$_\odot^{-1}$, which is smaller than the one inferred for the Galactic field (${\sim10^{-4}}$~CVs~M$_\odot^{-1}$).
This efficient destruction of primordial CV progenitors is somewhat consistent with the role of dynamics expected from observations.
For instance, 47 Tuc has a CV mass density of $\sim7\times10^{-5}$~CVs~M$_\odot^{-1}$, i.e. smaller than in the Milky Way field.
In addition, Cheng et al. (2018a~\cite{Cheng_2018}) showed that, unlike what was previously thought and claimed, the faint X-ray populations, primarily composed of CVs and ABs, are under-abundant in GCs with respect to the Solar neighbourhood and Local Group dwarf elliptical galaxies.

%
%
Regarding the dynamical formation of CVs in GCs, Belloni et al. (2019~\cite{Belloni_2019}) found that, even though strong dynamical interactions are able to trigger CV formation in binaries that otherwise would never become CVs, the detectable CV population is predominantly composed of CVs formed via typical common-envelope evolution ($\gtrsim 70$ per cent).
In particular, these authors showed that the main dynamical scenarios proposed in the literature have a very low probability of occurring, which resulted in (or very weak, if at all) correlation between the predicted number of detectable CVs and the predicted GC stellar encounter rate.
As discussed in Section~\ref{Sec2.3}, these finding are supported by the recent observations by Cheng et al. (2018a~\cite{Cheng_2018}), but in conflict with previous observational works.
Focusing now on bright and faint CVs, Belloni et al. (2019~\cite{Belloni_2019}) found that, on average, non-core-collapsed models tend to have small fractions of bright CVs, while core-collapsed models tend to have higher fractions.
In particular, most non-core-collapsed models have, on average, only $\sim7-33$~\%, which is consistent with that observed in 47~Tuc and $\omega$~Cen, while core-collapsed models have, on average, $\sim5-45$~\% bright CVs. The GCs in those models that are more similar to observed ones (i.e. compact, having high central surface brightness values, and short half-mass relaxation times) harbor even higher fractions of bright CVs ($\gtrsim50$~\%), which is consistent with the value observed in NGC~6397 and NGC~6752.
This supports the claim that, even though it is not expected to be dominant, dynamics plays a sufficiently important role to explain the relatively larger fraction of bright CVs observed in core-collapsed clusters.

%
%
Even though the GC outer parts are usually ignored in both observational and theoretical works, Belloni et al. (2019~\cite{Belloni_2019}) investigated CV populations in this region and found similar results to those obtained by Davies (1997~\cite{Davies_1997}). Davies found that a considerable fraction of pre-CVs may be formed sufficiently far from the GC central parts (i.e. far from the deleterious effects of a crowded environment), and eventually evolve to form CVs, especially in less evolved clusters.
Belloni et al. (2019~\cite{Belloni_2019}) found that, on average, a large fraction ($\lesssim50$~\%) of the entire population of predicted detectable CVs are expected to reside outside the half-light radii, and proposed that future observational efforts could be put towards the search for CVs in regions not close to the cluster centres.
This result is consistent with how we expect mass segregation to operate in stars clusters, i.e. (i) the greater the CV mass, the stronger the mass segregation, and (ii) the shorter the cluster's half-mass relaxation time, the stronger the mass segregation.
That said, it is not surprising that GCs with relatively long half-mass relaxation times (longer than a few Gyr) could have a considerable amount of CVs in the outer parts and the opposite for clusters with shorter relaxation times (shorter than a few hundred Myr).
These authors then suggest that in GCs having relatively long half-mass relaxation times, the number of CVs identified could significantly increase when looking for them in the outer parts.
Interestingly, spectra of such CVs could be relatively easy to obtain, since crowding should not be a big problem in the outer parts.
With respect to observations, Cheng et al. (2019a~\cite{Cheng_2019a}; 2019b~\cite{Cheng_2019b}; 2020~\cite{Cheng_2020}) recently investigated the phenomenon of mass segregation in the GCs 47~Tuc, Terzan~5 and M28.
These authors confirmed that there are some X-ray sources located outside the half-light radii and showed that the radial distribution of X-ray sources in these clusters is bimodal, with features similar to those found in the radial distribution of blue stragglers, they being modulated by mass segregation of primordial binaries.

%
%
Belloni et al. (2019~\cite{Belloni_2019}) also proposed an explanation for the spatial distribution of bright and faint CVs, dynamically formed or not, in core-collapsed and non-core-collapsed GCs.
These authors suggested that the properties of bright and faint CVs could be explained by means of the pre-CV and CV formation rates, their properties at their formation times and cluster relaxation times.
For non-dynamical CVs, most WD--MS binaries (i.e. pre-CVs) that are progenitors of detectable CVs are formed before $\sim1$~Gyr and they take $\gtrsim9$~Gyr to become CVs. This allows them to have enough time to sink to the central parts, provided the half-mass relaxation time is sufficiently short. The fact that bright CVs are younger and more massive than faint CVs makes them, in general, more centrally concentrated, as observed in NGC~6397 and NGC~6752. On the other hand, if bright and faint CVs have similar total masses, then they should have similar spatial distributions, as seen in 47 Tuc.
For pre-CVs dynamically formed in the core, these authors pointed out that they cannot stay in the core after their formation, since the formation process is very energetic and such pre-CVs will be likely expelled far from the core (see the fig.~11 in Belloni et al. 2019~\cite{Belloni_2019}). If the cluster half-mass relaxation time is short enough (as in core-collapsed GCs) to allow the pre-CV to segregate before it becomes a CV, such a population can come back to the central parts, thus forming a bright CV in the central parts, which is likely the case for at least some bright CVs in NGC~6397 and NGC~6752. However, for GCs having longer half-mass relaxation times, those bright CVs that are dynamically formed will be found far from the central parts, which might be the case for some of the bright CVs found near the half-light radii of 47~Tuc and $\omega$~Cen.

%
%
The last important finding by Belloni et al. (2019~\cite{Belloni_2019}) we address here is associated with the GC initial conditions.
As we have already mentioned, there is a huge degeneracy regarding GC modelling (i.e. different initial GC models evolve to comparable present-day global GC properties), which is one of the major problems in this sort of modelling.
This degeneracy could be broken, to some extent, by modelling properties of particular type of objects (e.g. CVs, binary black holes, blue stragglers, among others), which strongly depend on the environment properties in which they live.
In the particular case of CVs, Belloni et al. (2019~\cite{Belloni_2019}) found that GC models set with the Kroupa \ibp~and low CE efficiency ($\alpha_{\rm CE}\lesssim0.5$) better reproduce the observed amount of CVs in NGC~6397, NGC~6752, and 47~Tuc.
This is an important step forward and provides further support to the Kroupa \ibp, which has already been tested against both numerical simulations and observations and has successfully explained the observational features of young clusters, associations, and even binaries in old GCs (e.g. Belloni et al. 2017c~\cite{Belloni_2017c}, and references therein).
Moreover, within the framework of this \ibp, the radius--mass relation inferred by Marks \& Kroupa (2012~\cite{Marks_2012}) are in good agreement with the observed density of molecular cloud clumps, star-forming regions and GCs, and provide dynamical evolutionary time-scales for embedded clusters consistent with the life-time of ultra-compact H\,{\sc ii} regions and the time-scale needed for gas expulsion to be active in observed very young clusters, as based on their dynamical modelling (e.g. Belloni et al. 2018a~\cite{Belloni_2018a}, and references therein).

\section{Discussion}
\label{SecDISC}

We have briefly presented over the previous two sections the current status of our knowledge related to CVs in different types of GCs from a theoretical and an observational perspectives.
We discuss here the main issues in such results, which is intrinsically connected with employed observational approaches and interpretation of observational results.
In addition, we complement the discussion with interpretation of the observational results based on numerical simulation outcomes.

\subsection{On the nature of cataclysmic variables in globular clusters}
\label{Sec4.1}

As mentioned in Section~\ref{Sec2.1}, given the small number of DN outburst detections in the works by Shara et al, and Pietrukowicz et al., they concluded that DNe should be rare in GCs.
As already pointed out by Knigge (2012~\cite{Knigge_2012MMSAI}), both groups arrived to the same conclusion based on the properties of observed CVs in the Milky Way field, which seems to be a biased sample of the real population of CVs in the field.
In fact, if most CVs in the field are short-period CVs and period bouncers, then the observed CV population in the field (especially the bright ones) is not representative of the real population of CVs in the Galaxy.
Using distances from the \gaia~mission, Belloni et al. (2020a~\cite{Belloni_2020a}) showed that, despite of being relatively large, the homogeneous Galactic CV sample from the \sdss~is still biased towards bright systems.
The conclusion that DNe are rare in GCs based on a comparison with the Milky Way field does not seem necessarily correct, since it could just be that they are difficult to observe, especially in GCs, in which the CVs are expected to be an even older population (Belloni et al. 2017a~\cite{Belloni_2017a}).
Another very important point is related to the chosen filters and cadence.
For example, Pietrukowicz et al. (2008~\cite{Pietrukowicz_2008}) did not find any DNe in NGC~6752 during their campaign of 7 days in one year using V images, but Thomson et al. (2012~\cite{Thomson_2012}) found two DNe in outburst using NUV \hst~images, taken in a period of only $\sim3$~weeks. 
Another example comes from Modiano et al. 2020 \cite{Modiano_2020} who confirmed 3 DNe in 47 Tuc using $Swift$ UV images taken in different years. Despite they could not resolve the central parts of the cluster and the long cadence of their observations, their study also shows the influence of observational biases, which are important to overcome in order to make comparisons with models. 
As previously discussed, these examples demonstrate that the use of UV filters plays a fundamental role to find these elusive objects, as it is the band where DNe (and in general CVs) mainly emit.

As previously mentioned, a better way of interpreting this `dwarf nova problem' is to think of it as simply a lack of DN outbursts in GCs.
This could then lead to two main interpretations about the nature of GC CVs: either they are considerably old (i.e. systems with very small duty cycles) or they are essentially magnetic systems.
In what follows, we will argue that, instead of having only one of these cases, we mostly likely have a combination of them.

Regarding the first alternative, we discussed in Section~\ref{SecSIM} that recent numerical simulations predict very short DN duty cycles among the majority of GC CVs.
From the simulations by Belloni et al. (2019~\cite{Belloni_2019}), provided the typical CV mass transfer rates, most predicted detectable CV accretion discs would be unstable at regions far away from the WD surface.
This indicates that the strength of the WD magnetic field should be relatively strong in order to prevent disc instability and, in turn, outbursts in such systems.
In addition, most predicted faint CVs have extremely short duty cycles ($\lesssim6$~\%), which suggests that detecting outbursts amongst faint CVs is rather improbable.
On the other hand, these simulations also predict that part of the bright DNe would be recovered in multi-epoch searches, since most predicted bright CVs have duty cycles greater than $\sim10$~\%.
These recent numerical simulations, therefore, do not offer evidence for an overabundance of magnetic CVs in GCs, in comparison with the fraction found in the Milky Way.

Regarding the second alternative, from the fact that intermediate polars have disrupted accretion discs and polars are discless, the scenario in which the paucity of DN outbursts in GCs is explained by CVs being preferentially magnetic has gained strength in the last couple of decades (e.g. Dobrotka, Lasota \& Menou 2006~\cite{Dobrotka_2006}).
This would imply that GC CVs are hugely different from Milky Way CVs, since in only $\sim33 \%$ of known CVs, the WD magnetic field is sufficiently strong to control at least the inner part of the accretion flow and radiation properties (Pala et al. 2020~\cite{Pala_2020}).
Apart from the dearth of observed DN outbursts, other observational arguments supposedly supporting the idea of most GC CVs being magnetic are: (i) the detection of relatively strong He\,{\sc II} emission lines in some CVs in NGC~6397, which would be consistent with their magnetic nature (Grindlay et al. 1995~\cite{Grindlay_1995}); (ii) GC CVs look like Milky Way DNe in optical wavelengths and like intermediate polars in X-rays; (iii) the X-ray-to-optical ratios of GC CVs are much higher than those found among non-magnetic CVs in the Milky Way field.
These three points together suggest that GC CVs more likely resemble intermediate polars with low mass transfer rates.
In addition, Dobrotka, Lasota \& Menou (2006~\cite{Dobrotka_2006}) showed that CVs with low mass transfer rates and moderately strong WD magnetic moment do not easily exhibit DN-like ourbursts, which would be consistent with the paucity of detected DN outbursts and the observational properties mentioned above.

All these arguments have been discussed in detail by Knigge (2012~\cite{Knigge_2012MMSAI}), and here we just provide some general considerations.
Regarding point (i), it is important to highlight that not only magnetic CVs exhibit He\,{\sc II} emission lines in their spectra.
Nova-likes are also characterized by this feature and they are non-magnetic CVs. Interestingly, low-mass X-ray binaries also exhibit such characteristic (e.g. Van Paradijs \& Van der Klis 2001~\cite{VanParadijs_2001}, and references therein).
Concerning points (ii) and (iii), since the selection of CVs in GCs is strongly X-ray biased, that would automatically favour the detection of magnetic CVs.
This is because magnetic CVs, among all CV sub-groups, are the most luminous in X-rays, as the post-shock region (emitting region) can reach relatively very high temperatures, from a few up to dozens of keV.
So, it is likely that the CVs more luminous in X-rays are indeed magnetic CVs.
However, the same is not so easily claimed for the fainter systems.
Another point usually not considered is whether the production of magnetic CVs could be eventually enhanced in GCs.
Despite several scenarios have being proposed to explain the origin of magnetic WDs, especially in close binaries, there is currently no model that properly explains all observational constraints (e.g. Belloni et al. 2020b~\cite{Belloni_2020b}; Zorotovic et al. [in preparation], and references therein).
Therefore, it is not easy to infer whether magnetic CVs should be overproduced or not in GCs with respect to the Milky Way field.
As already mentioned, current models of CVs in GCs do not point towards an overabundance but further improvements in the modelling might shed light on this regard.

Despite the fact that the arguments towards magnetic CVs might be consistent and plausible, one should remember that our understanding about GC CVs are strongly biased towards luminous sources, as in the Milky Way field.
Even the faint populations described here so far are expected to be more luminous than the overwhelming majority of GC CVs. 
This suggests that this previously discussed claim could be reasonable mainly for the more luminous CVs that can be found in GCs nowadays.
Additionally, as pointed out by Pietrukowicz et al. (2008~\cite{Pietrukowicz_2008}), there is no real observational evidence that GC CVs are mostly magnetic and the best way to investigate whether the known CVs are magnetic or not is to search signatures of magnetic field such as strong and/or variable circular polarization and Zeeman splitting and/or cyclotron harmonics (humps) in their spectra.

Therefore, from the fact the most CVs should be DNe (Pala et al. 2020~\cite{Pala_2020}), from the intrinsic bias associated with either current technological limitations or observational approaches (Knigge 2012~\cite{Knigge_2012MMSAI}), from the absence of solid observational evidence that GC CVs are magnetic (Pietrukowicz et al. 2008~\cite{Pietrukowicz_2008}) and from the fact that we still do not understand the origin of the WD magnetic field in close binaries (e.g. Belloni et al. 2020b~\cite{Belloni_2020b}), the best alternative to the issue concerning the nature of GC CVs seems to be that 
 {\it many of the bright CVs (and perhaps some of the fainter systems), especially those with relatively large X-ray emission, could be magnetic. But the majority of the CV population in GCs are likely DNe, most of them being still undetected.}

\subsection{On the dynamical origin of most cataclysmic variables in globular clusters}
\label{Sec4.3}

In Section~\ref{Sec2.3}, we discussed a way to investigate connections
between the properties of X-ray sources with their host GC properties, by means of the correlation between the abundance of such sources with the GC stellar encounter rate $\Gamma$.
A correlation of this type is usually believed to indicate the dynamical origin of these sources.
Before proceeding further, we would like to highlight four important issues related to interpretation of this correlation, which are usually ignored.

%
%
First, $\Gamma$ depends on GC properties that might change quite substantially over the cluster life-time.
For instance, Hong et al. (2017~\cite{Hong_2017}, their fig.~1) showed that $\Gamma$ initially decreases rapidly with time, as the cluster expands in response to mass loss due to stellar evolution. 
After that, $\Gamma$ does not vary significantly.
However, if the cluster relaxation time is sufficiently short, $\Gamma$ slightly increases as the system evolves towards higher central densities and, eventually, core collapse.
When core contraction is halted (due to energy generated in the core), $\Gamma$ stops growing again.
After this first (deep) core collapse, which is associated with the largest value of $\Gamma$, the GC evolves having several core collapse episodes, which results, in turn, in several peaks of different strength in $\Gamma$. 
The time-scales for the several core-collapse episodes vary from cluster to cluster but are expected to be between a few Myr to a few Gyr, depending on the cluster properties.
Thus, having the present-day value of $\Gamma$ does not necessarily help in understanding the dynamical history of a GC.
This is specially true if one takes into account that different initial GC models might evolve to comparable present-day global GC properties (including $\Gamma$), which is a consequence of the huge degeneracy existent in GC modelling.
Therefore, addressing the impact of dynamics over the cluster life-time is a delicate issue which is not easily achieved with \textit{only} present-day global GC properties.

%
%
Second, from the observational point of view, these investigations are strongly X-ray biased, and it has been showed that not only contamination from background/foreground objects is common, but also from other types of compact binaries among faint X-ray sources. 
Therefore, in order to truly determine the influence of dynamics on CVs in GCs, complementary optical observations are necessary.  
For instance, from the 39 \chandra~X-ray sources that lie within the half-mass radius of NGC~6752, Lugger et al. (2017~\cite{Lugger_2017}) found 31 optical counterparts with \hst~imaging (20 newly identified by these authors).
Among them, there are 16 CVs (one probably a foreground system), nine ABs, three galaxies and three active galactic nuclei.
This means that $\sim15\%$ of the X-ray sources are not real cluster members and the majority of the identified faint X-ray sources seem to be CVs.
On the other hand, in NGC~6397, Cohn et al. (2010~\cite{Cohn_2010}) found 69 optical counterparts from 79 X-ray sources within its half-mass radius.
They correspond to 15 CVs, 42 ABs, two millisecond pulsars, one interacting galaxy pair and one  active galactic nucleus.
Thus, the majority of the faint X-ray sources in this cluster is composed of ABs (almost three times the amount of CVs), unlike NGC~6752. In these studies, both clusters have similar X-ray observing limits.
Therefore, having only X-ray information is clearly not enough to fully address the issue whether CVs (or at least the bright ones) are mostly dynamically formed or not.

%
%
Third, the number of X-ray systems detected is, as expected, very sensitive to the exposure time. Faint X-ray sources, mainly faint CVs and ABs, are the ones to be missed if exposure times are not long enough.
For example, in 47~Tuc, 108 sources were initially identified with \chandra~by Grindlay et al (2001~\cite{Grindlay_2001}).
With longer exposure time, Heinke et al. (2005~\cite{Heinke_2005}) identified 300 sources, and with even longer exposure time, Bhattacharya et al. (2017~\cite{Bhattacharya_2017}) identified 370 sources.
Similarly, in $\omega$~Cen, Haggard et al. (2009~\cite{Haggard_2009}) identified 180 sources in the first epoch of \chandra~data.
In a second epoch, with longer exposure time, Henleywillis et al. (2018~\cite{Henleywillis_2018}) identified 233 sources, 95 of which were new.

Fourth, another important aspect of this type of statistical studies is that they do not provide information about the history of the CVs, i.e., whether they should be predominantly primordial in origin or not.
This was already pointed out by Pooley \& Hut (2006~\cite{Pooley_2006}), who mentioned that their results would not be informative for any individual cluster alone.
Therefore, while these studies are extremely useful and valuable, further research, likely involving other wavelengths, has to be performed to fully assess the dynamical origin of CVs in GCs (Rivera Sandoval et al. [in preparation]).

Regarding the studies by Pooley \& Hut (2006~\cite{Pooley_2006}) and Cheng et al. (2018a~\cite{Cheng_2018}), it is important to first highlight the main differences in both approaches.
%
%
Pooley \& Hut (2006~\cite{Pooley_2006}) estimated the number of X-ray sources within the cluster half-mass radius by counting all detected sources and subtracting the estimated number of background sources based on the $\log\,N-\log\,S$ relationship of Giacconi et al. (2001~\cite{Giacconi_2001}).
However, this method is susceptible to dramatic changes with exposure time, as mentioned above and the number of background sources among the detected sources can also be large. 
For example, in cases like 47~Tuc, more than 70 background sources would be expected out of the 370 found so far (Heinke et al. 2005~\cite{Heinke_2005}; Bhattacharya et al. 2017~\cite{Bhattacharya_2017}). 
%
%
On the other hand, Cheng et al. (2018a~\cite{Cheng_2018}) used a simple method for studying the GC X-ray population, which is based on simply extracting all X-ray flux within the cluster half-mass radius.
In comparison with identifying individual X-ray sources, this is a much simpler method (see Heinke et al. 2020 \cite{Heinke_2020}, for a discussion regarding caveats in the approach by Cheng et al. 2018a~\cite{Cheng_2018}).
One problem, for example, is that when used on distant clusters or clusters with very few X-ray sources, the approach by Cheng et al. (2018a~\cite{Cheng_2018}) will be much less precise than for closer clusters. However, for their correlations, these authors removed clusters with low signal-to-noise ($<3$) and they dealt with the influence of background sources by removing them in a statistical way.

It is also important to note the relevance of the size of the sample when drawing conclusions.
For example, when considering a larger sample than that of Pooley \& Hut (2006~\cite{Pooley_2006}), as in the work by Cheng et al. (2018a~\cite{Cheng_2018}), the correlation between the abundance of X-ray sources and the GC stellar encounter rate does not seem to be as strong as found by the former authors.
This can be observed not only between these two works, but also between Pooley \& Hut (2006~\cite{Pooley_2006}) and Pooley (2010~\cite{Pooley_2010}).
Also, the fact that Cheng et al. (2018a~\cite{Cheng_2018}) reproduced the correlation for the sample analysed by Pooley \& Hut (2006~\cite{Pooley_2006}) suggests that, even though the weaker correlation in Cheng et al. when using their full sample could be due to the cluster larger distances and poorer quality of observations, one cannot ignore that, most likely, there is a bias due to small-number statistics, which is important to consider when interpreting such a correlation.

If we focus now only on the brightest CVs, one could think that the results by Pooley \& Hut (2006~\cite{Pooley_2006}) provides bulletproof observational evidence for the dynamical formation channel of the majority of these systems.
Indeed, several CVs with {${L_X>10^{31}}$~erg~s$^{-1}$} could have been produced dynamically.
However, it is important to emphasize that not every (spectroscopically) hard
\,\footnote{~We can separate X-ray sources into \textit{soft} and \textit{hard}, based on their properties as seen in their X-ray spectra. Soft systems have their emission dominated by low-energy photons, while hard systems have their emission dominated by high-energy photons.} 
CV with such luminosities can be attributed to dynamics, since, for example, intermediate polars in the Milky Way field are usually luminous, hard X-ray sources and optically bright (e.g. Mukai 2017~\cite{Mukai_2017}; Suleimanov et al. 2019~\cite{Suleimanov_2019})
In addition, the population II in Pooley \& Hut (2006~\cite{Pooley_2006}), claimed to be mostly composed of CVs, has systems with {${L_X>4\times10^{31}}$~erg~s$^{-1}$}.
While such a cut-off would encompass most of the bright CVs in the two core-collapsed GC discussed here, suggesting then some dynamical enhancement, most CVs in the two non-core-collapsed GCs would be left out.
Furthermore, not always an X-ray bright CV is also a bright optical or UV source. Indeed, in 47~Tuc, several optically faint CVs are strong X-ray emitters (Rivera Sandoval et al. 2018~\cite{Rivera_2018}).
These arguments then suggest that the CV correlation in Pooley \& Hut (2006~\cite{Pooley_2006}) is not always valid, and one should be cautious to consider it as a general probe into the dynamical formation of CVs in all types of GCs (Rivera Sandoval et al., [in preparation]).

So far we have showed that \textit{there are several issues involved in the studies related to faint X-ray sources and $\Gamma$, and that complementing such studies with a multi-wavelength approach would lead to a better understanding of the dynamical formation of CVs in GCs}. 
Such multi-wavelength approach is necessary in order to ``clean'' (by classifying) the existing samples of faint X-ray sources in GCs.
Not only with respect to membership, but also with respect to other types of sources that are not CVs, such as ABs. 
However, that is not always possible due to the large distances and high extinction towards many clusters, which prevents UV studies.
But in those cases, at least carrying out optical and X-ray studies will be more accurate than X-ray studies alone.
For instance, Cohn et al. (2010~\cite{Cohn_2010}) and Lugger et al. (2017~\cite{Lugger_2017}) showed that it is relatively easy to separate CVs from ABs using the X-ray to optical flux ratio, the one of CVs substantially exceeding those of ABs.

\section{Summary and conclusions}
\label{SecCONCL}

With the growth of available photometric, spectroscopic and X-ray GC data in the last couple of decades, we have been able to investigate, to some extent, several types of `exotic' objects  and, in particular, CV populations, in great detail in such environments.
The large amount of multi-wavelength observations carried out with \hst~and \chandra, and spectroscopy with \muse, on some well-studied GCs have been providing several constraints, which are rather useful for GC modelling.
Coupled with that, the development of theory, including numerical methods and binary evolution models, have allowed significant progress over more than three decades, aiming to explain the continuously updated observational constraints.

%
%
Despite accreting white dwarfs, such as CVs, are the most common expected class of compact interacting binaries to exist in GCs, a large number of these systems remain elusive from observations.
Indeed, recent numerical simulations predict that only $\sim2-4$~\% of the CV population is likely to be detected with current instrumentation. 
Thus, observational selection effects likely play a major role in the amount and properties of detected GC CVs.
For instance, the current samples are largely X-ray biased, as they are first detected as X-ray sources, which allow us to identify, amongst the whole population, only the systems that are more luminous in X-rays.
Moreover, a lot of faint CVs have been detected only through ultraviolet observations, instead of optical. 
Therefore, a very important step forward would be the attempt to reduce the bias in the current samples, by uniformly surveying several clusters using similar filters, especially near/far-ultraviolet ones, which are able to reveal many of the faint CVs.

%
%
One apparently problematic fact, likely connected with observational biases and known as the `dwarf nova problem', is the clear paucity of dwarf nova outbursts in GCs.
So far, only a few CVs have been identified through variability during outbursts in optical searches, either using ground-based or space-based data.
We discussed that this deficit of dwarf nova outbursts is unlikely explained by a large predominance of magnetic CVs in the GC populations, and is most likely associated with the intrinsically extremely short duty cycles of the majority of the population.
Indeed, most predicted CVs have very short dwarf nova outbursts in comparison with their recurrence time-scales, which implies very short duty cycle ($\lesssim6$~per~cent) and, in turn, extremely small probability of detection during outbursts.
Even though searching for variability could be a more effective technique with the adequate filters and cadences, given the current resources, it has been proven that a much more powerful technique is the combination of multi-wavelength photometry with \hst~and X-ray data with \chandra. 
These combined approaches have revealed a relatively large sample of CVs currently composed of more than a hundred systems distributed in several GCs.
We should also emphasize that the best way to verify whether most faint/bright CVs are indeed magnetic is to search for signatures of WD magnetic field in these systems, such as strong and/or variable circular polarization and Zeeman splitting and/or cyclotron harmonics (humps) in their spectra.

%
%
Amongst the most important findings based on the combination of \hst~and \chandra~in the last decade, we can mention the distinction between the CV populations in core-collapsed and non-core-collapsed GCs.
While the former seem to harbor two different populations, one brighter than the other, the latter seem to harbor only one population mostly composed of faint CVs. 
This distinction is also evident with respect to their spatial distribution. 
In core-collapsed clusters, the bright CVs are much more centrally concentrated than the faint ones, while in non-core-collapsed clusters, there is evidence for this only in the inner parts.
These properties of bright and faint CVs in different clusters can be understood by means of the pre-CV and CV formation rates, their properties at their formation times and the cluster relaxation times.
Most pre-CVs (i.e. WD--MS binaries) are expected to form before $\sim1$~Gyr of cluster evolution, and they take $\gtrsim9$~Gyr to become CVs (i.e. start mass transfer), which allows them to have enough time to sink to the central parts.
For clusters with relatively short relaxation times, regardless the formation channel, the fact that bright CVs are younger and more massive than faint CVs makes them, in general, more centrally concentrated, as observed in NGC~6397 and NGC~6752. 
However, if bright and faint CVs have similar average total masses, then they tend to have similar spatial distributions, as seen in 47~Tuc.
Nevertheless, further observational investigations of this type, including several other clusters, will potentially shed light on the processes of mass segregation operating in star clusters.

%
%
Another interesting observational fact is that the number of bright CVs per cluster mass in core-collapsed clusters is so far much higher than in non-core-collapsed clusters.
While in core-collapsed GCs with very short $T_{\rm rel}$ the observed fractions of bright CVs is in the range of $\sim40-60$~\%, in non-core-collapsed ones it is only $\sim20-30$~\%.
These results suggest that the formation of CVs is slightly favoured through strong dynamical interactions in core-collapsed GCs, due to the high stellar densities in their cores.
From recent numerical simulations, on average, non-core-collapsed models tend to have small fractions of bright CVs, while core-collapsed models tend have higher fractions.
These simulations then further suggest that dynamics plays a sufficiently important role to explain the relatively larger fraction of bright CVs observed in core-collapsed clusters.

%
%
Unlike what has been claimed for a long time, recent X-ray works related to the stellar encounter rate in GCs suggest that dynamics have a much less influence in the creation of CVs than previously believed. 
Previous works based on the detection of X-ray point sources in GCs  have found a strong correlation between the number of faint X-ray sources (mostly composed of CVs and active binaries) with the GC stellar encounter rate, while recent ones based on the X-ray emissivity, i.e. total X-ray luminosity per unit mass, suggest that this correlation, if existent, is rather weak.
One important problem in both approaches (if not the major one) is the contamination from other types of objects, since X-ray data alone does not seem enough to distinguish between the several types of faint X-ray emitters.
Therefore, the next required step, in this sort of analysis, is to have cleaner samples, in the sense of trying to unambiguously classify the objects, which is relatively easy, to some extent, with somewhat homogeneous photometric data.
From recent numerical simulations, it seems that dynamics more likely has a stronger influence in destroying CVs progenitors than in producing them.
In addition, even though strong dynamical interactions are able to trigger CV formation in binaries that otherwise would never become CVs, the detectable CV population is, on average, predominantly composed of CVs formed via typical common-envelope evolution ($\gtrsim 70$ per cent).
Nevertheless, this current observational conflict clearly shows that such a topic is not closed yet, and further observational investigations should be carried out to disentangle the issue whether there is or not a correlation between the number of faint X-ray sources, and CVs in particular, with the GC stellar encounter rate.

We end this document by mentioning that, in order to progress further with comparisons between numerical simulation outcomes and observations, it is important to derive properties such as orbital period, mass transfer rate and mass ratio for the observed GC CVs.
This seems possible with the analysis of data from several epochs of a substantial number of GCs. Data from \textit{HST} is fundamental given its unique capabilities of obtaining UV images and spectra with an exquisite spatial resolution that allows to resolve the crowded cores of GCs. Its larger field of view and larger sensitivity compared to ground-based imagers that are equipped with adaptive optics technology allows to study more extensive areas and detect much fainter CVs (and many other exotic binaries) in GCs. Therefore it is urgent to exploit its unique capabilities for the study of GCs before it stops operating. 
Another promising instrument is MUSE, which has demonstrated to have significant potential to confirm/identify CVs and other compact binaries in GCs.

\section*{Acknowledgements}

DB in debt to Mirek Giersz for valuables discussions on globular clusters that have been occurring for several years. DB would like to thank Joanna Miko{\l}ajewska, Krystian I{\l}kiewicz and Michael Shara for fruitful discussions on observations of cataclysmic variables and related objects in globular clusters during this workshop. DB would like to also thank Matthias R. Schreiber, M\'onica Zorotovic, Christian Knigge, J{\'o}zef Smak, Arkadiusz Olech and Claudia V. Rodrigues, for valuable discussions about cataclysmic variables in a more general perspective. DB was supported by the grant \#2017/14289-3, S\~ao Paulo Research Foundation (FAPESP).
LERS would like to thank to Maureen van den Berg, Rudy Wijnands, Thomas Maccarone and Craig Heinke for sharing their knowledge on compact interacting binaries in GCs and for the many valuable discussions specifically related to CVs. She is also grateful to many other close collaborators with whom she has exchanged interesting conversations about CVs in GCs. LERS acknowledges NASA for support under grant
80NSSC17K0334.


\bibliographystyle{JHEP}
\bibliography{references}


\bigskip
\bigskip

\noindent {\bf DISCUSSION}

\bigskip
\noindent {\bf ANNA F. PALA:} 
We see a lack of period bouncers in the Galactic-field CV population. Can this also be the reason for the lack of dwarf novae in GCs? Did you include period bouncers in your models?

\bigskip
\noindent {\bf DIOGO BELLONI:} 
Period bouncers are included in the models by Belloni et al. (2016~\cite{Belloni_2016a}; 2017a~\cite{Belloni_2017a}; 2017b~\cite{Belloni_2017b}; 2019~\cite{Belloni_2019}). One interesting difference between GCs and the Galactic disc is connected with the history of the star formation rate, which is rather different for both Galactic components. While in GCs we expect a/some burst(s) at the beginning, in the Galactic disc star/binary formation is expected to be roughly constant throughout its life-time, which makes GC CVs much older than the Galactic-field CV population. If the paucity of period bouncers in the Galactic CV population is somehow connected with longer evolutionary time-scales (e.g. Pala et al. 2020~\cite{Pala_2020}; Belloni et al. 2020a~\cite{Belloni_2020a}), we would have a higher fraction of period bouncers in GCs, but still a significant fraction of CVs around the orbital period minimum, which have very short duty cycles. Therefore, the apparent lack of dwarf novae in GCs more likely is a consequence of the intrinsically very short duty cycles of faint CVs, which require a considerably high amount of epoches in \hst~to recover their outbursts.

\bigskip
\noindent {\bf LINDA SCHMIDTOBREICK:} 
Do you find dependencies with the metallicity: evolution, duty cycle, etc?

\bigskip
\noindent {\bf DIOGO BELLONI:} 
All our models have \textit{only} one (low) metallicity ($Z=0.001$) and for this reason we could not address how GC CV properties are affected by this parameter. With respect to CV evolution, the dependence of CV properties on the metallicity is expected to be negligible (Stehle, Kolb \& Ritter 1997~\cite{Stehle_1997}). With respect to GC evolution, the metallicity affects the black hole mass spectrum, which in turn affects the global GC properties over the time-scale of the evolution. Even though, from model to model, metallicity will affect global properties, on statistical basis we do not expected that the picture presented here would be affected, especially taking into account that most CVs, on average, are expected to form through typical common-envelope events.

\bigskip
\noindent {\bf MICHAEL SHARA:} 
In $\sim25$~years of \hst~observations of 47~Tuc (47~epochs), we found only one more erupting dwarf nova; a very rarely erupting dwarf nova, supporting your results.

\end{document}